\documentclass[final,3p,times]{elsarticle}

\usepackage{geometry}
\usepackage{graphicx}
\usepackage{amssymb}
\usepackage{booktabs}

\usepackage{caption}
\usepackage{subfigure}
\usepackage{setspace}
\usepackage{lineno}
\usepackage{array}
\usepackage{float}
\usepackage{tcolorbox}
\usepackage{amsmath,amsthm}
\usepackage{mathrsfs}
\usepackage{algorithm, algorithmic}
\usepackage[colorlinks=true, allcolors=blue]{hyperref}

\usepackage[nodots]{numcompress}

\usepackage{arydshln}

\biboptions{sort&compress}
\usepackage[numbers]{natbib}

\theoremstyle{remark}

\journal{arXiv}

\begin{document}

\begin{frontmatter}

\title{Incremental Neural Controlled Differential Equations for Modeling of Path-dependent Material Behavior}

\author[label1]{Yangzi He}
\author[label1]{Shabnam J. Semnani\corref{cor1}}
\ead{ssemnani@ucsd.edu}
\date{}
\address[label1]{Department of Structural Engineering, University of California, San Diego, United States}

\cortext[cor1]{Corresponding author}

\begin{abstract}
Data-driven surrogate modeling has emerged as a promising approach for reducing computational expenses of multiscale simulations. Recurrent Neural Network (RNN) is a common choice for modeling of path-dependent behavior. However, previous studies have shown that RNNs fail to make predictions that are consistent with perturbation in the input strain, leading to potential oscillations and lack of convergence when implemented within finite element simulations.
In this work, we leverage neural differential equations which have recently emerged to model time series in a continuous manner and show their robustness in modeling elasto-plastic path-dependent material behavior. We develop a new sequential model called \textit{Incremental Neural Controlled Differential Equation} (INCDE) for general time-variant dynamical systems, including path-dependent constitutive models. INCDE is formulated and analyzed in terms of stability and convergence. Surrogate models based on INCDE are subsequently trained and tested for J2 and Drucker-Prager plasticity. The surrogate models are implemented for material point simulations and boundary value problems solved using the finite element method with various cyclic and monotonic loading protocols to demonstrate the robustness, consistency and accuracy of the proposed approach. 
\end{abstract}


\end{frontmatter}

\section{Introduction}\label{sec:intro}
A major challenge associated with multiscale modeling of nonlinear history-dependent materials is high computational costs of micro-scale simulations, which need to be performed at every iteration of the macro-scale problem \cite{Bishara:2023wh,Semnani:2020tg,Choo:2021vt,Karapiperis:2021vl}. Recently, data-driven and machine learning based approaches have gained popularity for modeling of complex materials and structural systems across scales. Surrogate models (or metamodels) such as neural networks have demonstrated the potential to reduce computational costs of nonlinear multiscale problems by replacing the microscale simulations conducted on representative volume elements (RVEs). Once trained, a surrogate model can be implemented within macroscale simulations to directly output average stress in the lower scale problem given the macroscopic strain as input. Existing studies have shown that application of recurrent neural networks to substitute the fine-scale models in FE$^2$ multiscale simulations can reduce the computing time by two orders of magnitude \cite{ghavamian2019accelerating, Logarzo2021}.

As surrogate constitutive models, neural networks have been used to directly return stress, potential energy, and/or stiffness. Feed-forward Artificial Neural Networks (ANNs) were the first type of neural networks applied for this purpose \cite{furukawa1998implicit,liu2021review, qu2023meta}. ANN implementations typically use previous and current strain, prior stress, internal state variables, and other properties specified for materials of interest to predict stress and any other relevant material descriptors. Recently, sequential models such as Recurrent Neural Networks (RNNs) have been applied as constitutive models due to their superior performance in learning path-dependent material behavior \cite{he2023machine,mozaffar2019deep, gorji2020potential, abueidda2021deep, bonatti2022importance, wang2022deep}. A number of researchers have used ANNs \cite{ge2021computational, fuhg2021model}, RNNs \cite{ghavamian2019accelerating,Logarzo2021, Wu2020,Yu:2024vj,Vijayaraghavan:2023vx}, and Convolutional Neural Networks (CNNs) \cite{Aldakheel:2023ts,Ahmad:2023vy} to accelerate the microscale simulations in multi-scale finite element analyses. 

Physics-informed machine learning approaches have boosted various areas in computational mechanics in the recent years. Physics can be incorporated into the modeling process in various ways, for example, via PDE residuals in loss functions \cite{Raissi2019}, neural network based differential operators \cite{wang2021physics}, and automatically-differentiable full-order models \cite{pachalieva2022physics}. Very recently, studies on surrogate constitutive modeling have shown interest in physics-informed machine learning models. The most common approach for incorporating physics into machine learning consists of adding physics-based terms as constraints to loss functions, similar to physics-informed neural networks (PINNs) \cite{Raissi2019}. Examples of such regularization terms include thermodynamics principles \cite{He:2022wu}, governing equations \cite{Taneja:2023tj}, derivatives such as stiffness and stress power \cite{zhang2022physics}, and Karush–Kuhn–Tucker condition \cite{Wang:2023tq}. Another category of methods perform a few intermediate steps by neural networks instead of directly substituting the entire mapping from strain to stress with a surrogate model. Examples include predicting intermediate variables used for numerical schemes in thermodynamics \cite{Masi:2022uo}, determining coefficients for stress spectral representations \cite{Fuhg:2022tq}, and estimating Cholesky factors used in calculation of symmetric positive definite tangential stiffness matrices \cite{xu2021learning}. Multi-fidelity ANN models have also been proposed to leverage high-fidelity experimental data and low-fidelity synthetic data \cite{su2023multifidelity}.

RNNs are commonly used for modeling sequential data (e.g. time series), for example, evolution of the state of dynamical systems. However, RNNs work as discrete approximations, which is problematic for irregularly sampled or partially observed data. Neural differential equations have been developed recently as continuous alternatives to RNNs, which are more natural choices for modeling dynamical systems and time series data. 
The original Neural Ordinary Differential Equation (Neural ODE or NODE) was designed and implemented by Chen et al. \cite{chen2018neural}, who also developed an efficient adjoint method to perform backpropagation. Subsequent studies modified the original NODE to Augmented Neural Ordinary Differential Equation (ANODE) \cite{dupont2019augmented, Norcliffe:2020vb}, GRU-ODE \cite{de2019gru, Jordan:2021wa}, ODE-RNN \cite{Rubanova:2019wq}, stabilized NODE \cite{Linot:2023ww}, and Neural Controlled Differential Equation (Neural CDE or NCDE) \cite{kidger2020neural}. ANODE augmented the representation space of state variables in original NODE, whereas NCDE and GRU-ODE models aimed to overcome the limitation that NODE can only model dynamical systems without control signals. The details and applications of various NODE models are reviewed in Section \ref{sec:intro-NODE}. Regularization has also been considered to increase the speed of training. For example, Kelly et al. \cite{kelly2020learning} penalized the loss function with the norm of higher-order time derivatives of NODEs to achieve high accuracy when a large step is taken by the Runge-Kutta solver. A number of researchers have studied the effects of numerical ODE solvers and integration schemes on the performance of NODEs \cite{Zhu:2022tc,Ott:2020wk}.

Neural differential equations provide an elegant interface between two modeling paradigms, i.e. machine learning and differential equations, which allows leveraging well-developed tools in each field. Additionally, since their formulation is continuous in time, they are applicable to irregular time series and missing data \cite{kidger2020neural}, and are more robust against perturbations to the inputs \cite{Yan:2019vk}. While Neural ODE is considered the continuous counterpart of ResNets, Neural CDE is the continuous counterpart of RNN. Recently, NODE has been applied to different areas of scientific computing, e.g. fluid mechanics \cite{portwood2019turbulence, rojas2021reduced,ayed2019learning}, robotics \cite{meleshkova2021application}, ecology \cite{bonnaffe2021neural}, and  geophysics \cite{kiani2022neural}. So far, very few studies have applied Neural ODEs to solid mechanics problems \cite{tac2022data,Tac:2023wk,Jones:2022tj}. 
In solid mechanics, Neural ODEs have been utilized to approximate the derivatives of strain energy functional with respect to deformation invariants for anisotropic hyperelastic materials \cite{tac2022data}. This method was subsequently extended to finite viscoelasticity \cite{Tac:2023wk}. Jones et al. \cite{Jones:2022tj} enhanced the NODE formulation by adding invariants of strain and strain rate as input controlling sequences, leading to the Internal state variable NODE (ISV-NODE) formulation based on the classical Coleman-Gurtin internal state variable theory.

Despite the promise of NODE in modeling sequential data and time-variant systems, its potential in modeling of path-dependent material behavior has remained unexplored.  
It has been shown that RNN-based surrogate models learn the path-dependent material behavior as a function of discrete time series and lack consistency \cite{bonatti2022importance, he2023machine}. That is, the output stress depends on the strain increment size, and the stress increment does not approach zero as strain increment decreases to zero. Moreover, the response of boundary value problem simulations with GRU-based surrogate models can oscillate and show unrealistic softening behavior at the onset of loading \cite{he2023machine}. In this work, we develop a new methodology based on NODE family of models to address these issues. We propose a new sequential model called Incremental Neural Controlled Differential Equation (INCDE) based on modification of NCDE. Subsequently, we develop a stabilized formulation for INCDE, derive its numerical solution, and study its convergence behavior. Finally, stress predictors composed of an INCDE and a decoder are trained to learn J2 and Drucker-Prager plasticity using a refined random-walk-based training data generation strategy. Multiple non-trivial loading protocols are utilized to study the robustness, consistency, accuracy, and stability of the surrogate model. In order to evaluate the efficacy of the model in finite element analyses, the stress predictor is integrated into a nonlinear finite element solver for solving boundary value problems.

This paper is organized as follows. Section \ref{sec:intro-NODE} introduces the concept and formulation of NODE and its variants. Section \ref{sec:problem} formulates NODE-based surrogate models for constitutive modeling. In Section \ref{sec:INCDE}, we discuss the motivation to propose Incremental Neural Controlled Differential Equations, numerical solution of the new model, stability and convergence of INCDE, and architecture of the INCDE-based surrogate model. Section \ref{sec:ML} discusses the technical details of the dataset and training. Section \ref{sec:examples} presents numerical examples of material point level simulations and boundary value problems to validate the proposed methodology and test the performance of the INCDE-based model.

\section{Introduction to Neural ODEs}\label{sec:intro-NODE}
Neural Ordinary Differential Equations (Neural ODEs or NODEs) 	are a novel family of sequential models in machine learning \cite{chen2018neural}. Contrary to RNNs where the hidden state is a discrete time-dependent variable, Neural ODE models the evolution of a continuous hidden state described as an initial value problem
\begin{equation}\label{NeuralODE}
    \dot{\mathbf z} = \hat{\mathbf f}(t, \mathbf z; \boldsymbol \theta) \, , \quad \mathbf z(t_0)= \mathbf z_0 \, ,
\end{equation}
where $\dot{\mathbf z}$ represents the time derivative of the hidden state, $\mathbf z$, and $t$ denotes time. $\hat{\mathbf f}(\cdot)$ is a function continuous in $t$ and Lipschitz continuous in $\mathbf z$, and $\boldsymbol \theta$ parameterizes the model. $\hat{\mathbf f}(\cdot)$ can be approximated with a neural network, e.g., feed forward \cite{chen2018neural}, convolutional neural network \cite{Yan:2019vk}, or a gated recurrent unit \cite{de2019gru}. It is noted that the forward propagation process in a Neural ODE does not output the hidden state, but its rate. A Neural ODE initial value problem can be integrated numerically to solve for the hidden state as time series ${\mathbf z(t_i)}$ for $i = 1, 2, ...$ as
\begin{equation}\label{NODEintegral}
    \mathbf z(t_1) = \mathbf z(t_0) + \int_{t_0}^{t_1} \hat{\mathbf f}(t, \mathbf z; \boldsymbol \theta) \,  \text{d}t \approx \textrm{ODESolve}(\hat{\mathbf f}, t_0, t_1, \mathbf z_0) \, .
\end{equation}

Dupont et al. \citep{dupont2019augmented} proposed the augmented NODE as follows to enhance the learning space and fit functions which NODE cannot represent:
\begin{equation}\label{augmented_NODE}
    \dot{\mathbf y} = \hat{\mathbf f}(t, \mathbf y; \boldsymbol\theta) \, ,
\end{equation}
where $\mathbf y = [\mathbf z^T, \mathbf a^T]^T$ is the concatenation of the NODE hidden state $\mathbf z$ and an augmented state, $\mathbf a$. Unlike $\mathbf z$, $\mathbf a$ is initialized as $\mathbf a(0) = \mathbf 0$. The augmented NODE has been applied to an optimal control problem for soft robotics \cite{Kasaei:2023vy} and combined with the generative adversarial networks to recover missing data in incomplete time series \cite{Chang:2023tc}.

A fundamental drawback of the original form of NODE is that Equation (\ref{NeuralODE}) is limited to dynamical systems without control, since the input space only includes time and the current hidden state. As a result, only the initial condition affects the trajectory of $\mathbf z$ and there is no way to control the system. For example, NODE is not applicable to modeling forced vibrations since Equation (\ref{NeuralODE}) can only describe free vibrations and there is no mechanism to exert external forces to the model. To address this problem, researchers have developed variations
of NODE which can be generalized as 

\begin{equation}\label{ANODE}
    \dot{\mathbf z} = \hat{\mathbf f}(t, \mathbf z, \mathbf x; \boldsymbol \theta) \, ,
\end{equation}
in which the input space of $\hat{\mathbf f}(\cdot)$ is supplemented with a new input variable $\mathbf x$ to enhance the representation of the original form. Despite $\hat{\mathbf f}$ representing a neural network, Equation (\ref{ANODE}) is actually the standard form of a time-variant nonlinear dynamical system, in which $\mathbf z$ is the state variable and $\mathbf x$ is the input signal. Back to the vibration example, $\mathbf x$ can pass load at time $t$, making Equation (\ref{ANODE}) a standard form for a controlled nonlinear dynamical system. In this paper, this family of models is referred to as Enhanced NODE (ENODE). Examples of ENODEs include the GRU-ODE \cite{de2019gru} and ISV-NODE \cite{Jones:2022tj}.

To address the same issue, Kidger et al. \cite{kidger2020neural} developed Neural Controlled Differential Equations (NCDE). In lieu of expanding the dimension of the hidden state, NCDE modifies Equations (\ref{NeuralODE}) and (\ref{NODEintegral}), respectively, as
\begin{equation}\label{NCDE}
    \dot{\mathbf z} = \hat{\mathbf F}( \mathbf z; \boldsymbol \theta)\dot{\mathbf x} \, ,
\end{equation}
and
\begin{equation}\label{NCDEintegral}
    \mathbf z(t_1) = \mathbf z(t_0) + \int_{t_0}^{t_1} \hat{\mathbf F}(\mathbf z; \boldsymbol \theta) \, \text{d}\mathbf x(t) \, ,
\end{equation}
where $\mathbf x(t)$ is the control sequence vector as a function of time. Also, $\hat{\mathbf F}(\cdot)$ denotes a neural network returning a matrix. the matrix-vector product is integrated over time to update the hidden state. This modification allows an exterior signal to control the dynamical system. Additionally, Kidger et al. \cite{kidger2020neural} have shown that NCDEs are universal approximators and the range of functions approximated by ENODE is a subset of NCDE with the same neural network parameters. NCDEs are theoretically more flexible to fit output data; therefore, they are promising models for general continuous dynamics. Various types of neural network representations of $\hat{\mathbf F}(\cdot)$ have so far been implemented for NCDEs, e.g., graph neural networks for learning graph embeddings \cite{qin2023learning}, and attention mechanism \cite{Vaswani:2017ul,jhin2021attentive}. 

The original NCDE is usually controlled by a smooth series of $\mathbf x$, so the integral in Equation (\ref{NCDEintegral}) can be rewritten as
\begin{equation}\label{NCDEintegral_time}
    \mathbf z(t_1) = \mathbf z(t_0) + \int_{t_0}^{t_1} \hat{\mathbf F}( \mathbf z; \boldsymbol \theta) \dot{\mathbf x} \, \text{d}t \, .
\end{equation}
Equation (\ref{NCDEintegral_time}) is converted to a standard NODE integral by letting $\mathbf G(t, \mathbf z, \mathbf x; \boldsymbol \theta) :=\hat{\mathbf F}(\mathbf z; \boldsymbol \theta) \dot{\mathbf x}(t)$. However, most control sequences in practice are sampled as discrete time series; therefore, the time series must be interpolated before solving the ODE. Performance of common interpolation methods for the NCDE is investigated by Morrill et al. \cite{morrill2021neural}. The optimal interpolation method is an open question. A related discussion is provided by Jhin et al. \cite{jhin2022exit, jhin2023learnable}.

\section{Problem Formulation}\label{sec:problem}
In this section, we first formulate and discuss the NODE family for modeling material nonlinearity. The formulations of ENODE-based and NCDE-based material nonlinear problems are covered in Section \ref{sec:NODE-plasticity}. Subsequent analyses in Section \ref{sec:INCDE} demonstrate the drawbacks of these two models for physics of dynamical systems, in particular for history-dependent material behavior. This motivates us to propose a new model called the Incremental Neural Controlled Differential Equation (INCDE). The formulation, numerical solution, properties, and architecture of the INCDE model are discussed in Sections \ref{sec:incde-formulation} to \ref{sec:incde-architecture}. For details of the finite element formulation, rate forms of J2 and Drucker-Prager plasticity, and their numerical solutions, we refer to \cite{he2023machine}.

\subsection{NODE-based Plasticity Modeling}\label{sec:NODE-plasticity} 

To alleviate the existing issues of RNN-based surrogate models discussed in Section \ref{sec:intro}, we consider Neural ODE family of models in this work. In the context of path-dependent material behavior, $\mathbf z$ mimics the internal state variables of plasticity models. Compared to the original NODE, ENODE and NCDE are more reasonable options for modeling path-dependent plasticity because strain and material properties, as external input signals, affect the dynamics of internal state variables. The ENODE-based surrogate model in this paper is expressed as
\begin{equation}\label{stressPredictor}
    \dot{\mathbf z} = \hat{\mathbf f}(\mathbf z, \boldsymbol \epsilon; \boldsymbol \theta_{\textrm{ENODE}}) \, ,
\end{equation}
which can be solved for a series of $\mathbf z$ using
\begin{equation}\label{ANODE_det_z}
	\mathbf z(t_1) = \mathbf z(t_0) + \int_{t_0}^{t_1} \hat{\mathbf f}(\mathbf z, \boldsymbol \epsilon; \boldsymbol \theta_{\textrm{ENODE}}) dt \approx \textrm{ODESolve}(\hat{\mathbf f}, t_0, t_1, \mathbf z_0, \boldsymbol \epsilon_0) \, .
\end{equation}
From the perspective of nonlinear dynamical systems, $\mathbf z$ is the state variable and $\boldsymbol\epsilon$ is the input strain signal. Subsequently, another neural network function $\hat{\mathbf g}(\cdot)$ is used to map $\mathbf z$ and $\boldsymbol{\epsilon}$ to stress, $\boldsymbol \sigma$, as
\begin{equation}\label{Decoder}
    \boldsymbol\sigma = \hat{\mathbf g}(\mathbf z, \boldsymbol\epsilon; \boldsymbol \theta_g) \, .
\end{equation}
Compared to the standard ENODE form in Equation (\ref{ANODE}), time $t$ in Equation (\ref{stressPredictor}) is omitted since it is not an explicit variable in rate-independent plasticity.

To develop the NCDE-based surrogate model, we use strain series as the control sequence and obtain
\begin{equation}\label{NCDEhardening}
    \mathbf z(t_1) = \mathbf z(t_0) + \int_{t_0}^{t_1} \hat{\mathbf F}(\mathbf z; \boldsymbol \theta_{\textrm{NCDE}}) \, \text{d}\boldsymbol\epsilon(t) \, ,
\end{equation}
in which $\mathbf z$ and $\hat{\mathbf F}(\cdot)$ resemble the the plastic internal variable and hardening modulus, respectively. This NCDE form also needs an additional decoder as in Equation (\ref{Decoder}) to map $\mathbf z$ to stress. The integral in Equation (\ref{NCDEhardening}) can be converted to the form of Equation (\ref{NCDEintegral_time}) as
\begin{equation}\label{NCDEstrain}
	\mathbf z(t_1) = \mathbf z(t_0) + \int_{t_0}^{t_1} \hat{\mathbf F}( \mathbf z; \boldsymbol \theta_{\textrm{NCDE}}) \dot{\boldsymbol\epsilon} \, \text{d}t \, .
\end{equation}

\section{Incremental Neural Controlled Differential Equations}\label{sec:INCDE}
\subsection{Motivation}\label{sec:incde-motivation}
In this subsection, we discuss the limitations of NODE-based models in capturing path-dependent material behavior. First, the ENODE-based model of Equation (\ref{stressPredictor}) is not numerically consistent with strain perturbations. That is, predictions do not converge as strain increments decrease. To demonstrate this, consider the solution of Equation (\ref{ANODE_det_z})  using the forward Euler method
\begin{equation}\label{midpoint_original}
    \mathbf z^{<n+1>} = \mathbf z^{<n>} + \Delta t^{<n>} \hat{\mathbf f}(\mathbf z^{<n>}, \boldsymbol \epsilon^{<n>}) \, , 
\end{equation}
in which the superscript $<n>$ denotes the numerical solution at $t^{<n>}$.
$\Delta t^{<n>}$ is the nominal time increment to increase loading steps. In  Equation (\ref{midpoint_original}), when $\boldsymbol{\epsilon}^{<n+1>}\to\boldsymbol{\epsilon}^{<n>}$, $\mathbf{z}^{<n+1>}$ does not necessarily approach $\mathbf{z}^{<n>}$.
This is because $\Delta t^{<n>}$ is the only factor which can determine self-consistency, but it does not approach zero. 
On the contrary, the NCDE form is numerically consistent with the change in strain. In Equation (\ref{NCDEhardening}), since $\text{d}\boldsymbol\epsilon$ is multiplied by the neural network output, $\text{d}\mathbf z$ vanishes if $\text{d}\boldsymbol\epsilon\to \mathbf 0$. However, the NCDE-based model also has potential challenges during implementation as discussed below. 

\begin{figure}[ht]
    \centering
    \includegraphics[width=0.5\textwidth]{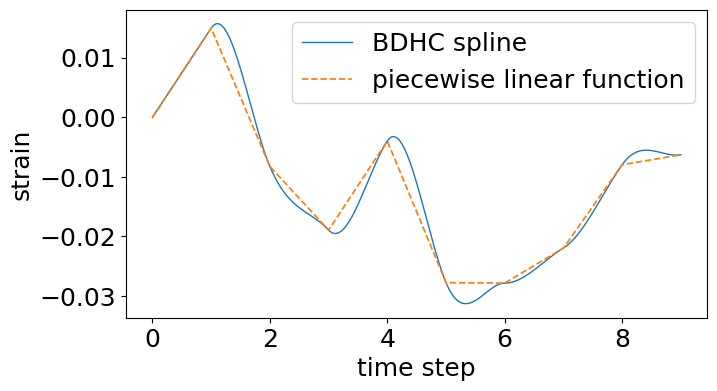}
    \caption{Example of an interpolated time series using backward Hermite cubic splines and piecewise linear functions.}
    \label{bhcs}
\end{figure}

Numerical solution of the NCDE-based model is obtained using Equation (\ref{NCDEstrain}), which requires evaluation of $\dot{\boldsymbol\epsilon}$. Since the strain series in finite element analyses are discrete and non-smooth, they have to be interpolated to obtain $\dot{\boldsymbol\epsilon}$. A critical requirement of strain interpolation is to meet causality: in finite element analysis, when strain at step $n+1$ is used to determine stress, the solver cannot rely on the strain in steps after $n+1$ to perform interpolation. This requirement rules out splines with derivative continuity conditions on two endpoints at steps $n$ and $n+1$.
Possible interpolators include piecewise linear functions and the backward difference Hermitian cubic (BDHC) splines \cite{morrill2021neural}. BDHC splines interpolate each interval $(\boldsymbol\epsilon^{<n>}, \boldsymbol\epsilon^{<n+1>})$ using a cubic spline which preserves $C^1$ continuity at two endpoints, where the derivative continuity is satisfied approximately with backward differences. Accordingly, only points before the interval are used and the causality requirement is met. Figure \ref{bhcs} shows an example of a strain series interpolated using piecewise linear and BDHC interpolation.

The BDHC spline interpolation has two major disadvantages. First, it drastically increases the computational expenses and memory usage of surrogate models. This cubic spline stores four coefficients per strain component, resulting in high complexity in time and space. Moreover, the interpolation is necessary for every new input series, making the NCDE-based stress predictor cumbersome to deploy. Second, this method affects the strain path. As illustrated in Figure \ref{bhcs}, the cubic spline overestimates the complexity of the path, which essentially changes the strain history and consequently the path-dependent material response. Compared to the cubic spline, linear interpolation makes weaker assumptions and minimizes the impact of interpolation on the path-dependent problem.

\subsection{INCDE Formulation}\label{sec:incde-formulation}
To address the problems described in Section \ref{sec:incde-motivation}, we modify the original NCDE form to develop Incremental Neural Controlled Differential Equations (INCDE). 
Strain and the rate of strain are added to the input of function $\hat{\mathbf F}$ to reflect path-dependency, leading to the rate form
\begin{equation}\label{initial_incde}
    \dot{\mathbf z} = \hat{\mathbf F}(\mathbf z, \boldsymbol \epsilon, \dot{\boldsymbol{\epsilon}}; \boldsymbol \theta_{\textrm{INCDE}})\dot{\boldsymbol{\epsilon}} \, .
\end{equation}
It is noted that $\dot{\boldsymbol{\epsilon}}$ included as input of function $\hat{\mathbf F}$ is not a control signal. Instead, it is a dependent variable approximated by the strain series and explicitly included as input to $\hat{\mathbf F}$. The integrals from Equations (\ref{NCDEhardening}) and (\ref{NCDEstrain}) are modified to obtain the hidden state update formula for INCDE in incremental form as
\begin{equation}\label{eq:incde-int}
    \begin{split}
        \mathbf z^{<n+1>} &= \mathbf z^{<n>} + \int_{\boldsymbol{\epsilon}^{<n>}}^{\boldsymbol{\epsilon}^{<n+1>}} \hat{\mathbf F}(\mathbf z, \boldsymbol\epsilon, \dot{\boldsymbol\epsilon}; \boldsymbol \theta_{\textrm{INCDE}}) \, \text{d}\boldsymbol\epsilon\\
        &\approx \mathbf z^{<n>} + \int_{0}^{1} \hat{\mathbf F}(\mathbf z, \boldsymbol{\epsilon}^{<n>} + t \Delta\boldsymbol\epsilon^{<n+1>}, \Delta\boldsymbol\epsilon^{<n+1>}; \boldsymbol \theta_{\textrm{INCDE}}) \Delta\boldsymbol\epsilon^{<n+1>} \, \text{d}t \, .
    \end{split}
\end{equation}
Unlike NCDE, the integral is now defined individually in the interval $(\boldsymbol{\epsilon}^{<n>}, \boldsymbol{\epsilon}^{<n+1>})$ for $n =1,2,..., N$, where $N$ is the size of the strain series. A nominal time variable $t\in[0, 1]$ is defined which allows piecewise approximation of $\dot{\boldsymbol\epsilon}$ as $\Delta\boldsymbol\epsilon^{<n+1>}:=\boldsymbol\epsilon^{<n+1>} - \boldsymbol\epsilon^{<n>}$. 
The INCDE model conserves the consistency of NCDE and alleviates the problems related to causality and strain rate calculation mentioned in Section \ref{sec:incde-motivation}, since input $\boldsymbol\epsilon^{<n+1>}$ is sufficient to determine the hidden state path. Since only $\boldsymbol{\epsilon}^{<n>}$ and $\boldsymbol{\epsilon}^{<n+1>}$ are needed to parameterize the integrand in Equation (\ref{eq:incde-int}) in an incremental fashion, the model is called \textit{Incremental Neural Controlled Differential Equation}.

\subsection{Stabilized INCDE}\label{sec:stabilize-incde}

Numerical integration corresponding to Equation (\ref{eq:incde-int}) can suffer from stability issues due to two main reasons. First, in finite element analysis, strain increments are generally non-uniform and so is the training dataset in this work, designed to augment the distribution of strain increments (see Section \ref{sec:data}). Therefore, the traditional step-size refinement for stabilizing explicit solvers is not applicable and solutions can easily blow up. Moreover, while strain series are normalized to $[-0.5, 0.5]$ (see Section \ref{sec:training}), the state variable $\mathbf{z}$ is not restricted and may grow exponentially through numerous integration steps. Large values of $\mathbf z$ dominate the input variables, leading to a positive feedback loop to $\mathbf z$ and instability of the numerical solution.

We alleviate this stability problem by integrating the state variable in a transformed space. For this purpose, the transformation $\mathbf Z = \tanh(\mathbf z)$ is introduced and we have
\begin{equation}\label{INCDE_diff_ztran}
    \text{d}\mathbf Z = \tanh'(\mathbf z) * \text{d}\mathbf z = (\mathbf 1 - \mathbf Z^2) * \text{d}\mathbf z = (\mathbf 1 - \mathbf Z^2) * \hat{\mathbf F}(\mathbf z, \boldsymbol\epsilon, \dot{\boldsymbol{\epsilon}}) \, \text{d}\boldsymbol\epsilon = \hat{\mathcal F}(\mathbf Z, \boldsymbol\epsilon, \dot{\boldsymbol{\epsilon}}) \, \text{d}\boldsymbol\epsilon \, ,
\end{equation}
in which ``$*$" denotes element-by-element multiplication for matrices or vectors. Hidden state $\mathbf z$ is replaced with the transformed state $\mathbf Z$ using
\begin{equation}
    \text{d}\mathbf z = (\mathbf 1 - \mathbf Z^2)^{-1} * \hat{\mathcal F}(\mathbf Z, \boldsymbol\epsilon, \dot{\boldsymbol{\epsilon}}) \, \text{d}\boldsymbol\epsilon \,.
\end{equation}
This form is unfavorable in numerical analysis since $(\mathbf 1 - \mathbf Z^2)^{-1}$ has singularities as any components $Z_{i} \to \pm 1$. Instead, let 
\begin{equation}
    \hat{\mathcal N}(\mathbf Z, \boldsymbol\epsilon, \dot{\boldsymbol{\epsilon}}) \, \text{d}\boldsymbol\epsilon = (\mathbf 1 - \mathbf Z^2)^{-1} * \hat{\mathcal F}(\mathbf Z, \boldsymbol\epsilon, \dot{\boldsymbol{\epsilon}}) \, \text{d}\boldsymbol\epsilon \, ,
\end{equation}
and assume that the limit of $\hat{\mathcal N}$ exists as any element in $\mathbf Z $ approaches $\pm 1$. Therefore, a modified differential form is obtained as
\begin{equation}\label{stable_differential_form}
    \text{d}\mathbf Z = (\mathbf 1 - \mathbf Z^2) * \hat{\mathcal N}(\mathbf Z, \boldsymbol\epsilon, \dot{\boldsymbol{\epsilon}}) \, \text{d}\boldsymbol\epsilon \, ,
\end{equation}
which can be integrated to obtain $\mathbf Z$. This stability modification is valid if $\mathbf Z$ is bounded in range $(-1, 1)$ after integration. In fact, we show that boundedness of $\hat{\mathcal N}$ is a sufficient condition for this. If $\hat{\mathcal N}$ has bounded output, a differential $\text{d}\mathbf w$ can then be defined as $\text{d}\mathbf w = \hat{\mathcal N} \,\text{d}\boldsymbol\epsilon$. The differential form in Equation (\ref{stable_differential_form}) is rewritten as
\begin{equation}\label{eq:dw}
     \text{d}\mathbf w =  (\mathbf 1 - \mathbf Z^2)^{-1} * \text{d}\mathbf Z \, .
\end{equation}
Integrating Equation (\ref{eq:dw}) and letting the initial value $\mathbf Z^{<0>} = \boldsymbol{0}$ yields
\begin{equation}
    \mathbf Z =  \tanh(\mathbf w) \, ,
\end{equation}
thus all components of $\mathbf Z$ are bounded in range $(-1, 1)$.

The multiplier $(\mathbf 1 - \mathbf Z^2)$ serves as a \textit{self-damping factor} since it renders $\text{d}\mathbf Z\to \mathbf 0$ as $\mathbf Z\to\pm \mathbf 1$ and changes the dynamics of the system. In addition, $\mathbf z$ no longer appears in the new differential form and is omitted from the formulation. $\hat{\mathcal N}$ is the neural network model which is trained for predicting the transformed hidden state $\mathbf Z$, to be used as input of the stress decoder. Since the stabilized INCDE form of Equation (\ref{stable_differential_form}) is used in all following sections, it is simply referred to as INCDE.
A straightforward method to guarantee stability is to append a bounded activation function (e.g., tanh) as the final layer of $\hat{\mathcal N}$. For the numerical solution of an INCDE, the stability condition holds with proper strain increment sizes to meet the assumption that the strain increment is a differential in the theoretical form. Stability is also illustrated numerically in Section \ref{sec:example1}.

\subsection{Numerical Solution to INCDE}\label{sec:numerical-incde}

\subsubsection{Numerical Integration}\label{sec:incde-integarl}
INCDE model of Equation (\ref{stable_differential_form}) is expressed in rate form as
\begin{equation}\label{stab_INCDE}
    \dot{\mathbf Z} = (\mathbf 1 - \mathbf Z^2) * \hat{\mathcal N}(\mathbf Z, \boldsymbol\epsilon, \dot{\boldsymbol\epsilon}) \,\dot{\boldsymbol\epsilon}\\ \,.
\end{equation}
Incremental form of INCDE is based on piecewise linear interpolation of strain in each interval. Accordingly, strain is expressed as $\boldsymbol\epsilon = \boldsymbol{\epsilon}^{<n>} + t\Delta\boldsymbol{\epsilon}^{<n+1>}$, where $t\in [0, 1]$ is a time variable defined in every interval formed by two strain steps. Also, strain rate at any time between steps $n$ and $n+1$ is a constant determined using $\dot{\boldsymbol{\epsilon}} = (\boldsymbol{\epsilon}^{<n+1>} - \boldsymbol{\epsilon}^{<n>}) / 1 = \Delta\boldsymbol\epsilon^{<n+1>}$. Substituting the incremental form of strain and strain rate into Equation (\ref{stab_INCDE}) gives the differential form in interval $(\boldsymbol\epsilon^{<n>}, \boldsymbol{\epsilon}^{<n+1>})$ as
\begin{equation}\label{univariate_form}
    \dot{\mathbf Z} = (\mathbf 1 - \mathbf Z^2)*\hat{\mathcal N}(\mathbf Z, \boldsymbol\epsilon^{<n>} + t\Delta\boldsymbol\epsilon^{<n+1>}, \Delta\boldsymbol\epsilon^{<n+1>})\Delta\boldsymbol\epsilon^{<n+1>},\ t\in[0, 1] \, .
\end{equation}
Equation (\ref{univariate_form}) introduces the standard multivariate initial value problem
\begin{equation}\label{init_val_problem}
    \dot{\mathbf Z} = \hat{\mathbf N}^{<n+1>}(\mathbf Z, t) \, , \, \mathbf Z_{0} = \mathbf Z^{<n>} \, , \, t\in[0, 1] \, ,
\end{equation}
where $\hat{\mathbf N}^{<n+1>}(\mathbf Z, t) := (\mathbf 1 - \mathbf Z^2)*\hat{\mathcal N}(\mathbf Z, \boldsymbol\epsilon^{<n>} + t\Delta\boldsymbol\epsilon^{<n+1>}, \Delta\boldsymbol\epsilon^{<n+1>})\Delta\boldsymbol\epsilon^{<n+1>}$ and $\mathbf Z_0$ is the initial value. Standard ODE solvers such as the forward Euler method, the midpoint method, and the fourth-order Runge-Kutta (RK4) method can be implemented to solve for $\mathbf Z$. Implicit methods (e.g., backward Euler and Crank-Nicolson methods) are not recommended since they require solving nonlinear systems derived from neural networks, leading to high computational costs. For a time step between $t_1$ and $t_2$ in $[0, 1]$ with a uniform increment $\Delta t := t_2 - t_1$, the Euler, midpoint, and RK4 methods are implemented as follows.

Euler method:
\begin{equation}\label{euler}
        \mathbf Z^{<n+t_2>} = \mathbf Z^{<n+t_1>} + \Delta t \, \hat{\mathbf N}^{<n+1>}(\mathbf Z^{<n+t_1>}, t_1)
\end{equation}

Midpoint method:
\begin{equation}\label{midpoint}
    \begin{cases}
        \mathbf Z^{<n+t_1+\frac{\Delta t}{2}>} = \mathbf Z^{<n+t_1>} + \frac{\Delta t}{2} \, \hat{\mathbf N}^{<n+1>}(\mathbf Z^{<n+t_1>}, t_1)\\
        \mathbf Z^{<n+t_2>} = \mathbf Z^{<n+t_1>} + \Delta t \, \hat{\mathbf N}^{<n+1>}(\mathbf Z^{<n+t_1+\frac{\Delta t}{2}>}, t_1 + \frac{\Delta t}{2})
    \end{cases}
\end{equation}

RK4 method:
\begin{equation}\label{rk4}
    \begin{cases}
        \mathbf k_1 = \hat{\mathbf N}^{<n+1>}(\mathbf Z^{<n+t_1>}, t_1)\\
        \mathbf k_2 = \hat{\mathbf N}^{<n+1>}(\mathbf Z^{<n+t_1>} + \frac{\Delta t}{2} \, \mathbf k_1, t_1 + \frac{\Delta t}{2})\\
        \mathbf k_3 = \hat{\mathbf N}^{<n+1>}(\mathbf Z^{<n+t_1>} + \frac{\Delta t}{2} \, \mathbf k_2, t_1 + \frac{\Delta t}{2})\\
        \mathbf k_4 = \hat{\mathbf N}^{<n+1>}(\mathbf Z^{<n+t_1>} + \Delta t \, \mathbf k_3, t_1 + \Delta t)\\
        \mathbf Z^{<n+t_2>} = \mathbf Z^{<n+t_1>} + \frac{\Delta t}{6}(\mathbf k_1 + 2\mathbf k_2 + 2\mathbf k_3 + \mathbf k_4)
    \end{cases}
\end{equation}

\subsubsection{Convergence of Numerical Solution}\label{sec:convergence}
Although the initial value problem in Equation (\ref{init_val_problem}) is piecewise, the convergence rate of $\mathbf Z$ should still agree with the orders of truncation errors of Euler, midpoint, and RK4 methods. Since the ODE is locally continuous, assuming that $\mathbf Z^{<n>}$ has no error, the truncation error in $(\boldsymbol\epsilon^{<n>}, \boldsymbol{\epsilon}^{<n+1>})$ corresponding to the $\alpha$-th order numerical method is 
\begin{equation}\label{accum_err}
    E(\Delta\mathbf Z^{<n+1>}) := \Vert \Delta\mathbf Z^{<n+1>} - \Delta\mathbf Z^{*<n+1>} \Vert_2 \sim O(\Delta t^\alpha) \, ,
\end{equation}
where $\Delta\mathbf Z^{*<n+1>}$ is the exact increment of $\mathbf Z$. In the first $k$ intervals, the sum of all truncation errors $E(\cdot)$ determines the truncation error of $\mathbf Z$ at step $k$, so we have
\begin{equation}
    \begin{split}
        E(\mathbf Z^{<k>}) &= \Vert \mathbf Z^{<k>} - \mathbf Z^{*<k>} \Vert_2 = \left\Vert \left(\mathbf Z^{<0>} + \sum_{k}\Delta \mathbf Z^{<k>} \right) - \left(\mathbf Z^{<0>} + \sum_k\Delta\mathbf Z^{*<k>} \right) \right\Vert_2 \\
        &= \left\Vert \sum_{k}\left(\Delta \mathbf Z^{<k>} - \Delta\mathbf Z^{*<k>} \right) \right\Vert_2 \leq k\sum_{k} E(\Delta \mathbf Z^{<k>}) \sim kO(\Delta t^{\alpha}) \sim O(\Delta t^{\alpha}) \, ,
    \end{split}
\end{equation}
in which $\mathbf Z^{*<k>}$ is the exact solution at step $k$. Therefore, the convergence rate of $\mathbf Z$ with respect to $\Delta t$ is still $\alpha$, the order of the truncation error of the numerical ODE solver. This convergence rate is also valid for the predicted stress as long as the decoder is Lipschitz continuous for $\mathbf Z$. That is, for a given decoder $\hat{\mathbf g}(\cdot)$, a fixed strain $\boldsymbol{\epsilon}$, the numerical $\mathbf Z^{<n+1>}$ for any $n$, and any value $\mathbf Z^{*}$, there exists a constant $K\geq 0$ such that 
\begin{equation}
    \Vert \boldsymbol\sigma^{<n+1>} - \boldsymbol\sigma^{*} \Vert_2 = \Vert \hat{\mathbf g}(\mathbf Z^{<n+1>}, \boldsymbol{\epsilon}) - \hat{\mathbf g}(\mathbf Z^*, \boldsymbol{\epsilon}) \Vert_2 \leq K  \Vert \mathbf Z^{<n+1>} - \mathbf Z^{*} \Vert_2 \, .
\end{equation}
Since $\mathbf Z^*$ and $\boldsymbol{\sigma}^*$ are arbitrary, setting $\mathbf Z^*$ and $\boldsymbol{\sigma}^*$ to the converged hidden state and the stress at step $n+1$ leads to 
\begin{equation}
    E(\boldsymbol\sigma^{<n+1>}) \leq K E(\mathbf Z^{<n+1>}) \sim O(\Delta t^\alpha) \, .
\end{equation}

Next, we show that the convergence rate of INCDE with respect to $\Delta\boldsymbol{\epsilon}$ is linear. Consider two initial value problems as Equation (\ref{init_val_problem}) but parameterized with different strain increments $\Delta\boldsymbol{\epsilon}_1$ and $\Delta\boldsymbol{\epsilon}_2$. The updated $\mathbf Z$ vectors read
\begin{equation}\label{updated_z1_z2}
    \begin{cases}
        \mathbf Z_1^{<n+1>} = \mathbf Z^{<n>} + \int_0^1 (\mathbf 1 - \mathbf Z^2) * \hat{\mathcal N}(\mathbf Z, \boldsymbol{\epsilon}^{<n>} + t\Delta \boldsymbol\epsilon_1, \Delta \boldsymbol\epsilon_1)\Delta \boldsymbol\epsilon_1 \, \text{d}t\\
        \mathbf Z_2^{<n+1>} = \mathbf Z^{<n>} + \int_0^1 (\mathbf 1 - \mathbf Z^2) * \hat{\mathcal N}(\mathbf Z, \boldsymbol{\epsilon}^{<n>} + t\Delta \boldsymbol\epsilon_2, \Delta \boldsymbol\epsilon_2)\Delta \boldsymbol\epsilon_2 \, \text{d}t \, .
    \end{cases}
\end{equation}
As Equation (\ref{updated_z1_z2}) holds for all steps, the superscripts ``$^{<n>}$" and ``$^{<n+1>}$" are dropped for simplicity: 
\begin{equation}\label{updated_z1_z2_simplified}
    \begin{cases}
        \mathbf Z_1 = \mathbf Z_0 + \int_0^1 (\mathbf 1 - \mathbf Z^2) * \hat{\mathcal N}(\mathbf Z, \boldsymbol{\epsilon} + t\Delta \boldsymbol\epsilon_1, \Delta \boldsymbol\epsilon_1)\Delta \boldsymbol\epsilon_1 \, \text{d}t\\
        \mathbf Z_2 = \mathbf Z_0 + \int_0^1 (\mathbf 1 - \mathbf Z^2) * \hat{\mathcal N}(\mathbf Z, \boldsymbol{\epsilon} + t\Delta \boldsymbol\epsilon_2, \Delta \boldsymbol\epsilon_2)\Delta \boldsymbol\epsilon_2 \, \text{d}t \, .
    \end{cases}
\end{equation}
The difference $\delta\mathbf Z = \mathbf Z_2 - \mathbf Z_1$ due to the perturbation $\delta\boldsymbol\epsilon = \Delta\boldsymbol\epsilon_2 - \Delta\boldsymbol\epsilon_1$ is 
\begin{equation}\label{z_difference}
    \delta \mathbf Z := \int_0^1 (\mathbf 1 - \mathbf Z^2) * \left[ \hat{\mathcal N}(\mathbf Z, \boldsymbol{\epsilon} + t\Delta \boldsymbol\epsilon_2, \Delta \boldsymbol\epsilon_2)\Delta \boldsymbol\epsilon_2 - \hat{\mathcal N}(\mathbf Z, \boldsymbol{\epsilon} + t\Delta \boldsymbol\epsilon_1, \Delta \boldsymbol\epsilon_1)\Delta \boldsymbol\epsilon_1 \right] \, \text{d}t \, .
\end{equation}
Assuming that $\hat{\mathcal N}$ is first-order differentiable, we can write the Taylor expansion of $\hat{\mathcal N}(\mathbf Z, \boldsymbol{\epsilon} + t\Delta \boldsymbol\epsilon_2, \Delta \boldsymbol\epsilon_2)$ at $\Delta \boldsymbol\epsilon_1$ as
\begin{equation}\label{taylor}
    \hat{\mathcal N}(\mathbf Z, \boldsymbol{\epsilon} + t\Delta \boldsymbol\epsilon_2, \Delta \boldsymbol\epsilon_2) = \hat{\mathcal N}_1 + t\partial_{\boldsymbol\epsilon}\hat{\mathcal N}_1 \delta\boldsymbol{\epsilon} + \partial_{\Delta\boldsymbol{\epsilon}}\hat{\mathcal N}_1 \delta\boldsymbol{\epsilon} + \textrm{H.O.T.} \, ,
\end{equation}
in which $\hat{\mathcal N}_1$ denotes $\hat{\mathcal N}(\mathbf Z, \boldsymbol{\epsilon} + t\Delta \boldsymbol\epsilon_1, \Delta \boldsymbol\epsilon_1)$. Substituting Equation (\ref{taylor}) into Equation (\ref{z_difference}) yields

\begin{equation}\label{dz_vs_dep_intermediate_step}
    \begin{split}
    \delta \mathbf Z &= \int_0^1 (\mathbf 1 - \mathbf Z^2) * \left[ \hat{\mathcal N}_1 (\Delta\boldsymbol{\epsilon}_2 - \Delta\boldsymbol{\epsilon}_1) + t\partial_{\boldsymbol\epsilon}\hat{\mathcal N}_1 :(\Delta\boldsymbol{\epsilon}_2\otimes \delta\boldsymbol{\epsilon}) + \partial_{\Delta\boldsymbol{\epsilon}}\hat{\mathcal N}_1:(\Delta\boldsymbol{\epsilon}_2\otimes \delta\boldsymbol{\epsilon}) + \textrm{H.O.T.} \right] \, \text{d}t \\
    &\approx \int_0^1 (\mathbf 1 - \mathbf Z^2) * \left[ \hat{\mathcal N}_1 \delta\boldsymbol{\epsilon} + t\partial_{\boldsymbol\epsilon}\hat{\mathcal N}_1 :(\Delta\boldsymbol{\epsilon}_2\otimes \delta\boldsymbol{\epsilon}) + \partial_{\Delta\boldsymbol{\epsilon}}\hat{\mathcal N}_1:(\Delta\boldsymbol{\epsilon}_2\otimes \delta\boldsymbol{\epsilon}) \right] \, \text{d}t \, .
    \end{split}
\end{equation}
The integral in Equation (\ref{dz_vs_dep_intermediate_step}) is split to
\begin{equation}\label{dz_vs_dep_intermediate_step2}
    \begin{split}
        \delta \mathbf Z &= \mathbf I_1 + \mathbf I_2 + \mathbf I_3\\
        & = \int_0^1 (\mathbf 1 - \mathbf Z^2) * \hat{\mathcal N}_1 \delta\boldsymbol{\epsilon} \, \text{d}t + \int_0^1 (\mathbf 1 - \mathbf Z^2) *  t\partial_{\boldsymbol\epsilon}\hat{\mathcal N}_1 :(\Delta\boldsymbol{\epsilon}_2\otimes \delta\boldsymbol{\epsilon}) \, \text{d}t \\
        &+ \int_0^1 (\mathbf 1 - \mathbf Z^2) * \partial_{\Delta\boldsymbol{\epsilon}}\hat{\mathcal N}_1:(\Delta\boldsymbol{\epsilon}_2\otimes \delta\boldsymbol{\epsilon}) \, \text{d}t \, . 
    \end{split}
\end{equation}
Given a matrix $\mathbf A$ and a vector $\mathbf b$, the inequality $\Vert \mathbf A \mathbf b \Vert_p \leq \Vert \mathbf A \Vert_p \Vert \mathbf b \Vert_p$ holds for all $p$, which yields
\begin{equation}\label{triple_ineqs}
    \begin{split}
        \Vert \mathbf I_1 \Vert &\leq \left\Vert\int_0^1 (\mathbf 1 - \mathbf Z^2) * \hat{\mathcal N}_1  \, \text{d}t \right\Vert \Vert \delta\boldsymbol{\epsilon} \Vert = C_1 \Vert \delta\boldsymbol{\epsilon} \Vert\\
        \Vert\mathbf I_2\Vert &\leq \left\Vert\int_0^1 (\mathbf 1 - \mathbf Z^2) *  t\partial_{\boldsymbol\epsilon}\hat{\mathcal N}_1 \Delta\boldsymbol{\epsilon}_2  \, \text{dt} \right\Vert \Vert \delta\boldsymbol{\epsilon} \Vert = C_2 \Vert \delta\boldsymbol{\epsilon} \Vert\\
        \Vert\mathbf I_3\Vert &\leq \left\Vert\int_0^1 (\mathbf 1 - \mathbf Z^2) * \partial_{\Delta\boldsymbol{\epsilon}}\hat{\mathcal N}_1\Delta\boldsymbol{\epsilon}_2 \, \text{dt}\right\Vert \Vert \delta\boldsymbol{\epsilon} \Vert = C_3 \Vert \delta\boldsymbol{\epsilon} \Vert \, . 
    \end{split}
\end{equation}
Finally, the triangle inequality leads to 
\begin{equation}\label{dz_vs_dep_tri_ineq}
    \Vert\delta \mathbf Z\Vert = \Vert \mathbf I_1 + \mathbf I_2 + \mathbf I_3 \Vert \leq \Vert \mathbf I_1 \Vert + \Vert \mathbf I_2 \Vert + \Vert \mathbf I_3 \Vert \leq (C_1 + C_2 + C_3) \Vert\delta\boldsymbol{\epsilon}\Vert \, .
\end{equation}
Therefore, it is concluded that the truncation error is proportional to the norm of $\delta\boldsymbol{\epsilon}$, and the convergence rate with respect to strain increment is linear. Similarly, stress predictions also converge linearly with $\Delta\boldsymbol{\epsilon}$ if the decoder is Lipschitz continuous.

Finally, it is noteworthy that the analyses above show that $\mathbf Z$ and $\hat{\boldsymbol{\sigma}}$ can converge to a certain limit with the same convergence rate. However, this limit may not necessarily match the ground truth, and it is the main goal of INCDE training to map this numerical solution to the ground truth. In Section \ref{sec:ML}, some methods are discussed for this purpose. In addition, Section \ref{sec:example2} provides a stress prediction example at an integration point to study the effects of $\Delta t$ and different ODE solvers.

\subsection{Advantages of INCDE}\label{sec:incde-advantage}
The new INCDE formulation presented in this paper preserves the advantage of NCDE in maintaining consistency of the state variable with respect to the increment size of the control sequence, $\boldsymbol{\epsilon}$. The differences with NCDE include stabilization, augmentation of the input space, and piecewise formulation. The INCDE-based model is stabilized with a self-damping factor to prevent the state variables from exploding. The conditions for ensuring stability are readily met by neural networks. The controlling sequence, strain, and its rate are included as additional input variables, leading to a more flexible model for capturing history-dependent behavior.

\begin{figure}[ht]
    \centering
    \includegraphics[width=0.6\textwidth]{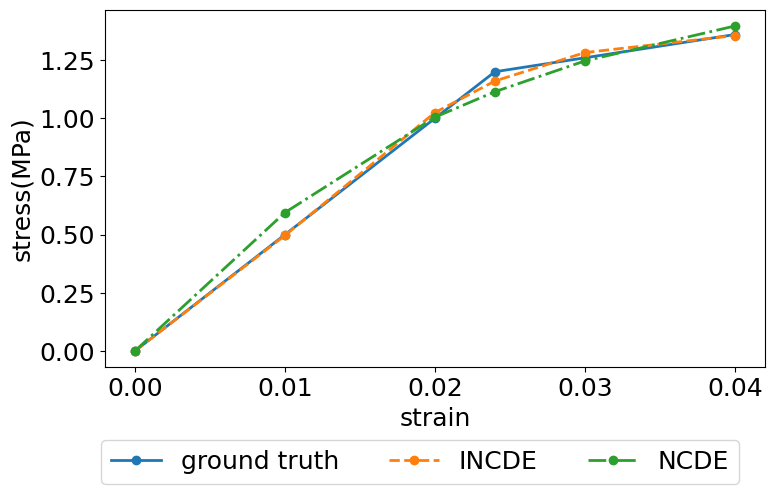}
    \caption{Comparison between INCDE and NCDE models for learning a 1D bilinear model of plasticity}
    \label{piecewise_ivp_conceptual_plot}
\end{figure}

From another perspective, piecewise functions provide better flexibility in capturing the response of physical systems compared to continuous analytic functions. For example, splines are generally better interpolators compared to $C^{\infty}$ analytic functions. Another example is the Galerkin method, which approximates physical field variables with a discretized space consisting of low-degree continuous functions.
Similarly, an INCDE-based model defines a piecewise initial value problem in every interval as expressed in Equation (\ref{init_val_problem}), where $\boldsymbol{\epsilon}^{<n>}$ and $\Delta\boldsymbol{\epsilon}^{<n+1>}$ in $\hat{\mathbf N}^{<n+1>}$ are perceived as piecewise coefficients parameterizing the ODEs. These piecewise initial value problems are capable of capturing functions with both low and high degrees of continuity.
To illustrate this property, consider the following one-dimensional INCDE and NCDE models, respectively
\begin{equation}\label{piecewise_example}
    \Delta \mathbf Z^{<n+1>} = \int_{0}^{1} (\mathbf \mathbf 1 - \mathbf Z^2)\hat{\mathcal N}(\mathbf Z, \epsilon^{<n>} + \, t\Delta\epsilon^{<n+1>}, \Delta\epsilon^{<n+1>}) \, \Delta\epsilon^{<n+1>} \, \text{d}t \, ,
\end{equation}
\begin{equation}\label{piecewise_example_ncde}
    \Delta \mathbf Z = \int_{t_1}^{t_2}\hat{\mathbf f}(\mathbf Z)\dot{\epsilon} \, \text{d}t \,. 
\end{equation}
The NCDE model calculates the strain rate using piecewise linear interpolation of strain. The decoder has a single bias-free linear layer such that $\sigma = \mathbf W \mathbf Z$. Both models are assigned feed-forward neural networks with 10 hidden states, 2 hidden layers with 20 neurons each, and nonlinear ELU() activation function. Equations (\ref{piecewise_example}) and (\ref{piecewise_example_ncde}) are solved using the midpoint method with $\Delta t = 1.0$. Models are trained for 1500 epochs in order to fit a 6-point bilinear model shown with the blue curve in Figure \ref{piecewise_ivp_conceptual_plot}.

It is observed from Figure \ref{piecewise_ivp_conceptual_plot} that the NCDE model underfits the strain data at all points, while INCDE has a smaller error at the yield point and fits all other points perfectly. Since NCDE is a global dynamical system defined over the entire time domain, it must reduce errors simultaneously at all points. In this problem, the most difficult task for the NCDE model is the abrupt transition upon yielding. INCDE, on the other hand, does not necessitate continuous global parameters since each interval is parameterized locally. Accordingly, the local error occurring at the transition point does not impact the model's quality in other intervals. 

\subsection{Architecture of INCDE Stress Predictor}\label{sec:incde-architecture}
\begin{figure}[ht]
    \centering
    \includegraphics[width=0.85\textwidth]{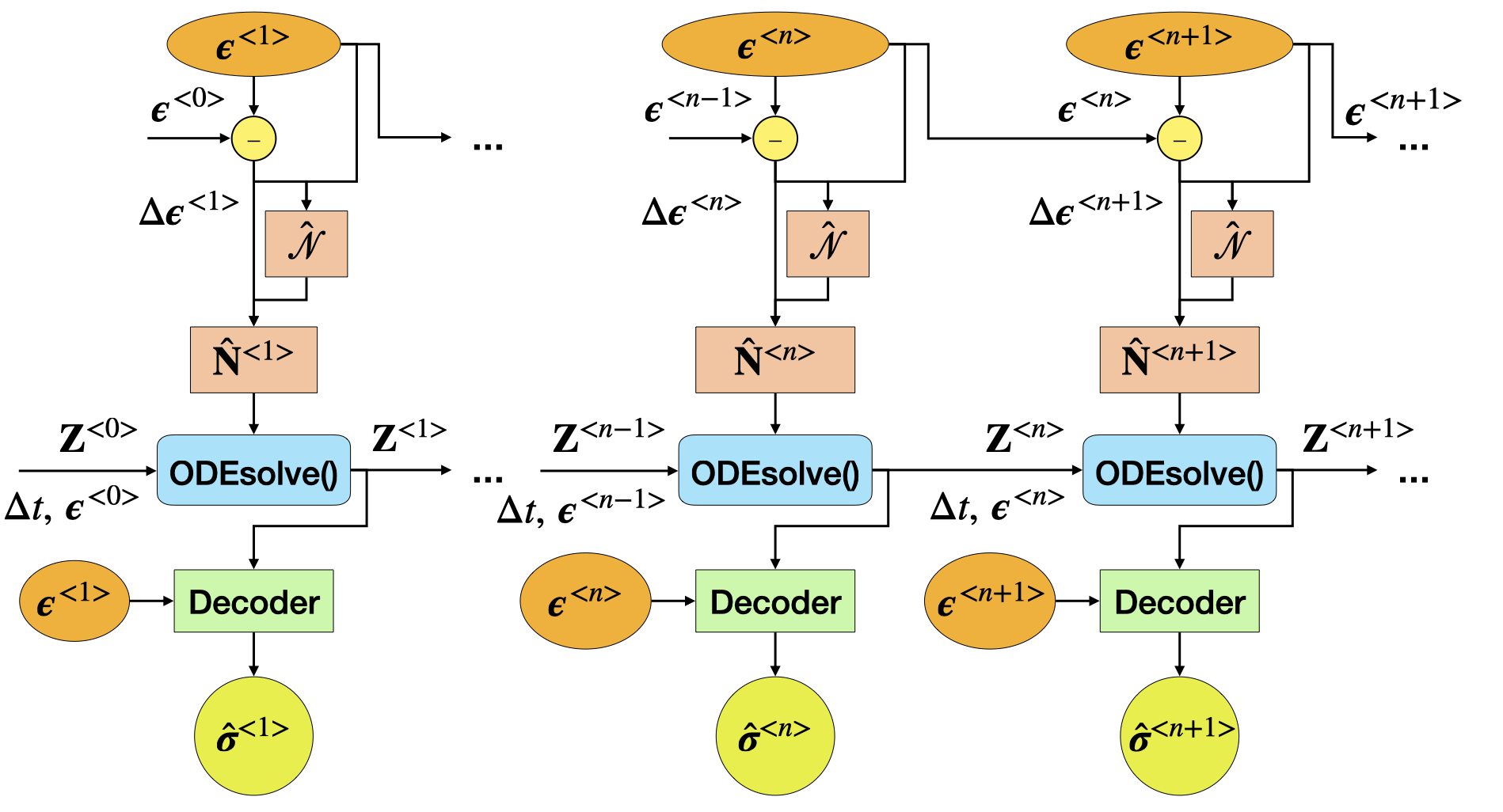}
    \caption{Architecture and forward propagation of the INCDE-based stress predictor. This graph shows evaluation of the INCDE model at steps $1$, $n$, and $n+1$. Strain $\boldsymbol{\epsilon}^{<n>}$ and hidden state $\mathbf Z^{<n>}$ propagate forward from step $n$ to $n+1$.}
    \label{architectures}
\end{figure}

Forward propagation of the comprehensive stress predictor is shown in Figure \ref{architectures}. At every step, new strain $\boldsymbol\epsilon^{<n+1>}$ is the external input. The previous strain $\boldsymbol\epsilon^{<n>}$ is subtracted from $\boldsymbol\epsilon^{<n+1>}$ to obtain the new strain increment $\Delta\boldsymbol\epsilon^{<n+1>}$ and update the parameters of $\hat{\mathcal N}$. The new initial value problem is then defined with $\hat{\mathbf N}^{<n+1>}$. Next, the initial value $\mathbf Z_0 = \mathbf Z^{<n>}$ and a specified nominal time step $\Delta t$ initialize the ODE solver and update the new hidden state $\mathbf Z^{<n+1>}$. $\mathbf Z^{<n+1>}$ is input to the decoder along with $\boldsymbol\epsilon^{<n+1>}$ to predict the new stress $\hat{\boldsymbol{\sigma}}^{<n+1>}$. Finally, $\mathbf Z^{<n+1>}$ propagates and initializes the next initial value problem. 

There are no requirements for network types of $\hat{\mathcal N}$ and the decoder, since the dynamics of the system is learned by the differential equation itself. Feed-forward neural networks are capable of learning $\hat{\mathcal N}$ and the decoder. Hyperparameters of the stress predictors in this work are listed in Table \ref{architecture_hyperparams}, which are tuned via grid search. Note that all layers in one network have the same number of hidden neurons during the grid search. For each combination of hyperparameters, the average minimum test loss from 5-fold cross-validation in the first 60 epochs is used as the selection metric.

\begin{table}[ht]
    \centering
    \caption{Hyperparameters of INCDE models trained for Drucker-Prager, J2 with isotropic hardening, and J2 with combined hardening.}
    \begin{tabular}{l l l l l l l}
        \hline
         & \multicolumn{2}{c}{\textbf{J2 (isotropic)}} & \multicolumn{2}{c}{\textbf{J2 (combined)}} & \multicolumn{2}{c}{\textbf{Drucker-Prager}}\\
        \textbf{Hyperparameter} & $\hat{\mathcal N}$ & Decoder & $\hat{\mathcal N}$ & Decoder & $\hat{\mathcal N}$ & Decoder\\[5pt]
        \hline 
         Number of hidden states in $\mathbf Z$ & 80 & -  & 80 & - & 80 & - \\[5pt]
         Number of input variables & $80 + 12$ & $80+6$ & $80 + 12$ & $80+6$ & $80 + 12$ & $80+6$ \\[5pt]
         Number of layers & 3 & 4 & 3 & 4 & 3 & 4\\[5pt]
         Number of hidden neurons & $125$ & $130$ & $125$ & $110$ & $120$ & $130$\\[5pt]
         Output size & $80\times6$ & 6 & $80\times6$ & 6 & $80\times6$ & 6 or 7\\[5pt]
         Intermediate Activation Functions & ELU() & ELU() & ELU() & ELU() & ELU() & ELU()\\[5pt]
         Output Activation Function & tanh() & - & tanh() & - & tanh() & - \\[5pt]
        \hline
    \end{tabular}
    \label{architecture_hyperparams}
\end{table}

There are a few important points in determining the architectures of $\hat{\mathcal N}$ and the decoder. For intermediate activation functions of $\hat{\mathcal N}$, a $C^{1}$ function is preferred. This choice meets the requirement for $\hat{\mathcal N}$ to be first-order differentiable to ensure the convergence of $\mathbf Z$ with respect to the strain increment $\Delta\boldsymbol\epsilon$, as discussed in Section \ref{sec:convergence}. For the decoder, all layers have zero bias terms to ensure zero strain and zero hidden state map to zero stress and avoid bias in stress predictions. In addition, all activation functions of the decoder need to be Lipschitz continuous, as discussed in Section \ref{sec:convergence}. Fortunately, most practical activation functions, including the ELU() function in this study, meet this requirement.

\begin{figure}[ht]
    \centering
    \includegraphics[width=0.35\textwidth]{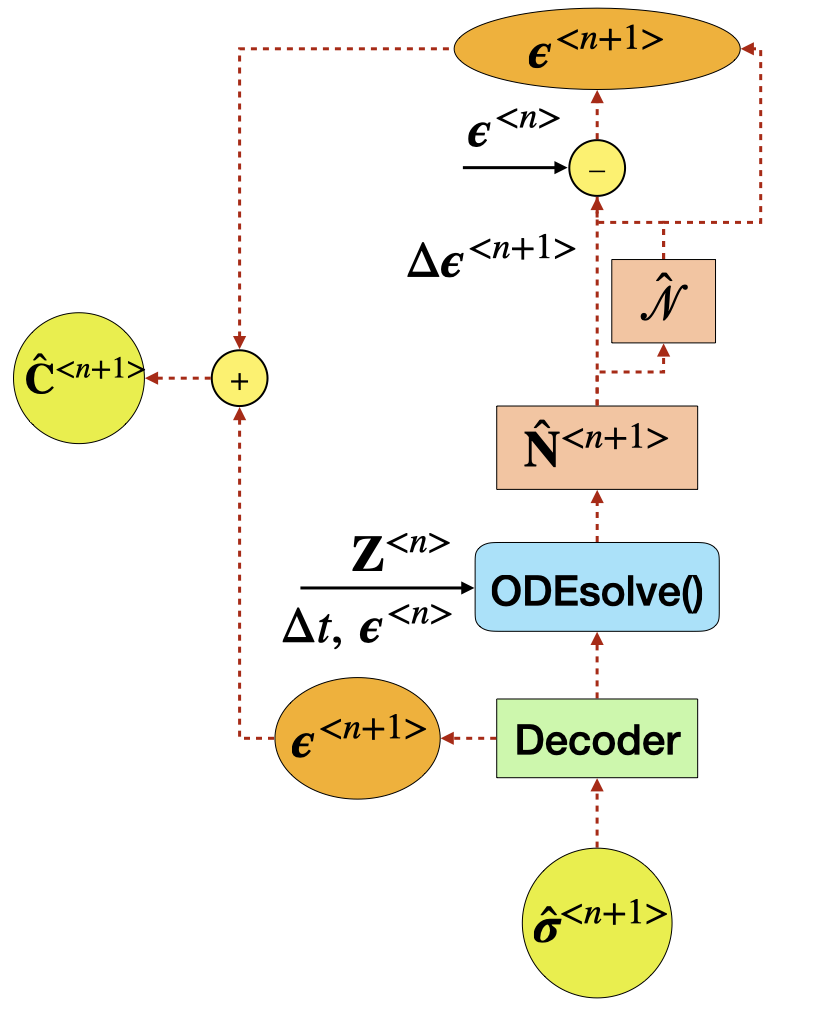}
    \caption{Computational graph for backpropagation of the INCDE-based network to determine the tangential stiffness tensor.}
    \label{INCDE_backprop}
\end{figure}

When INCDE is applied as a surrogate material model in finite element analysis, the backpropagation procedure is performed to calculate  the tangential stiffness $\hat{\mathbf C}^{<n+1>} := \partial\hat{\boldsymbol{\sigma}}^{<n+1>}/\partial\boldsymbol{\epsilon}^{<n+1>}$ to solve nonlinear systems. Figure \ref{INCDE_backprop} illustrates the backpropagation process using auto-differentiation. It is noted that $\Delta\boldsymbol\epsilon^{<n+1>}$ is a dependent variable of $\boldsymbol{\epsilon}^{<n+1>}$, so the flow of computation passes through the subtraction operator and reaches the top input through two different paths. For the INCDE model trained for Drucker-Prager which outputs pressure and deviatoric stress components (see Section \ref{sec:data}), backpropagation steps start with
\begin{equation}
    \partial\hat{\boldsymbol{\sigma}}^{<n+1>}/\partial\boldsymbol{\epsilon}^{<n+1>} = \partial \left(\hat p^{<n+1>}\mathbf 1 + \hat{\mathbf s}^{<n+1>}\right)/\partial\boldsymbol{\epsilon}^{<n+1>}
\end{equation}
in which $\hat p$, $\hat{\mathbf s}$, and $\mathbf 1$ are the predicted pressure, predicted deviatoric stress, and identity tensor expressed in Voigt notations. Subsequent steps remain the same as Figure \ref{INCDE_backprop}.

\section{Machine learning}\label{sec:ML}
\subsection{Data Acquisition}\label{sec:data}
A sufficiently representative dataset is critical for training a successful and generalizable machine learning model. To obtain a stress predictor for plasticity, the training dataset should be inclusive and cover potential patterns of input and output series. In this study, we use the random-walk-based data generation method developed by He and Semnani \cite{he2023machine}, since it is proven to be representative of possible stress and strain patterns. In this data generation approach, a strain series is created by drawing a random strain increment vector at each load step, which contains six random component magnitudes and directions (i.e., ``+" and ``-"). The component magnitudes and directions are drawn from a uniform distribution $U(0, 0.01)$ and a Bernoulli distribution $B(0.5)$, respectively. To keep a number of successive elastic loading/unloading steps which are common in engineering, the strain history is forced to remain in elastic phase over a randomly specified number of steps (see Algorithm 2 proposed in \cite{he2023machine}).

\begin{figure}[ht]
    \centering
    \subfigure[]{\includegraphics[width=7.7cm]{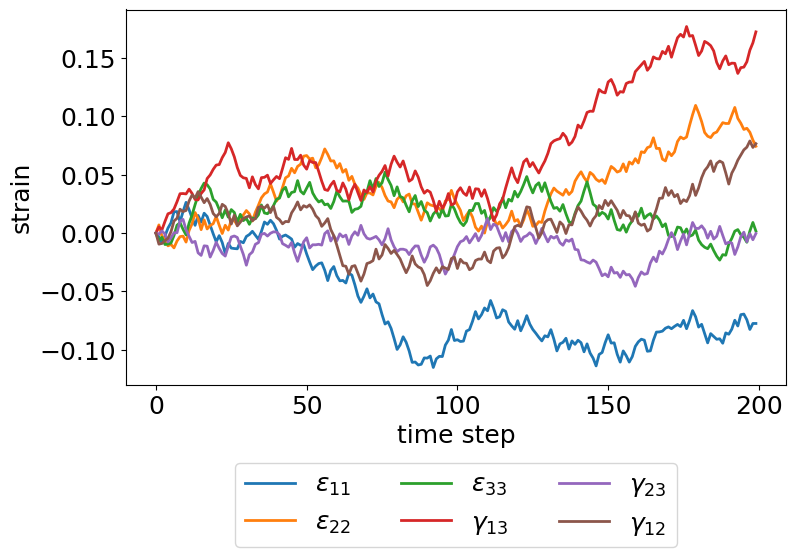}}
    \subfigure[]{\includegraphics[width=7.35cm]{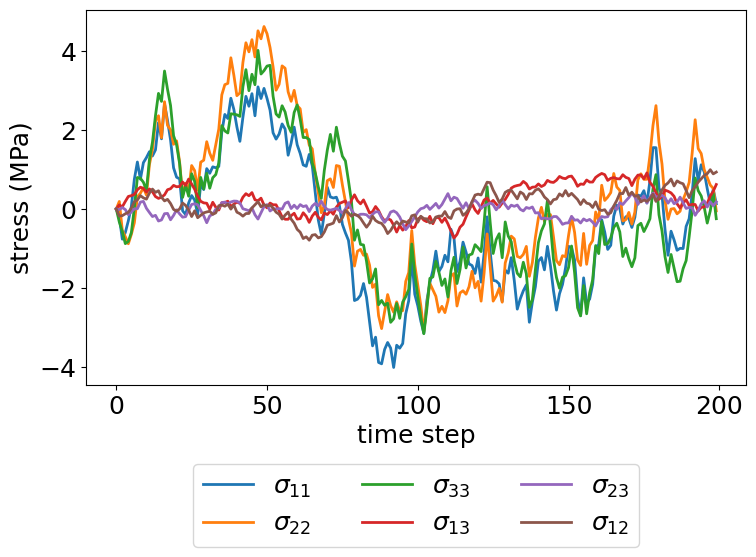}} 
    \subfigure[]{\includegraphics[width=7.7cm]{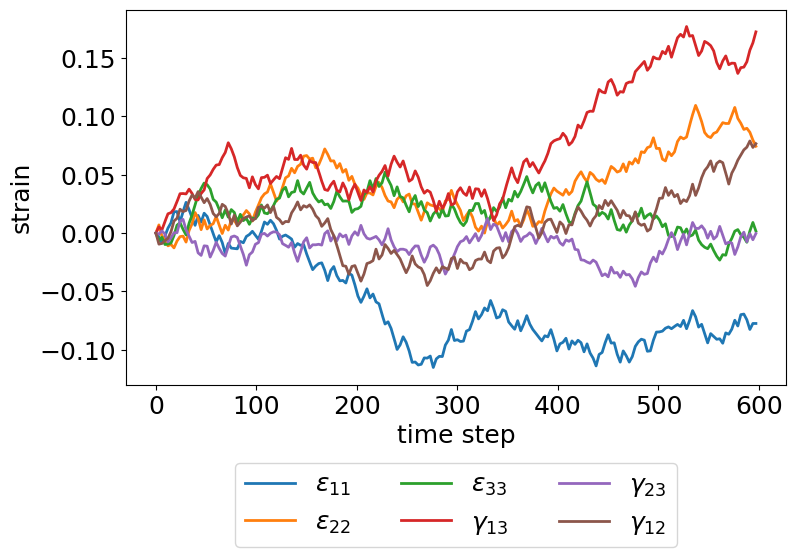}}
    \subfigure[]{\includegraphics[width=7.35cm]{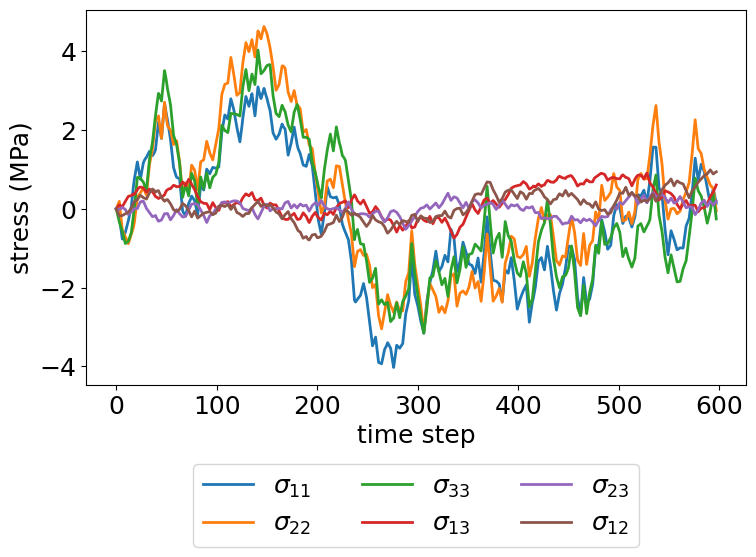}} 
    \caption{Strain and stress samples from the random-walk dataset before (a,b) and after (c,d) partitioning.}
    \label{training_data_sample}
\end{figure}

\begin{figure}[ht]
    \centering
    \subfigure[]{\includegraphics[trim =0 0cm 0 0.0cm, clip, width=7.4cm]{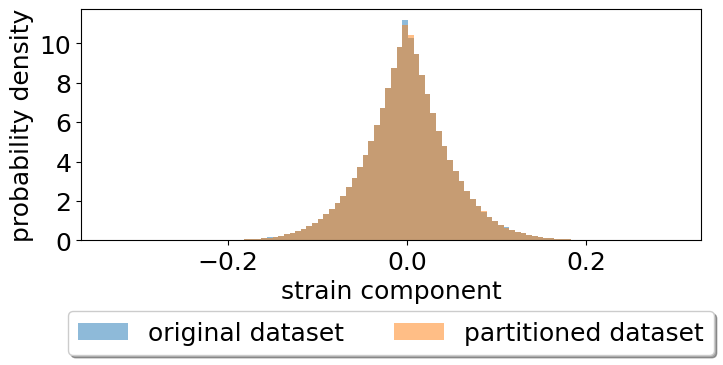} }
    \subfigure[]{\includegraphics[trim =0 0cm 0 0.0cm, clip, width=7.6cm]{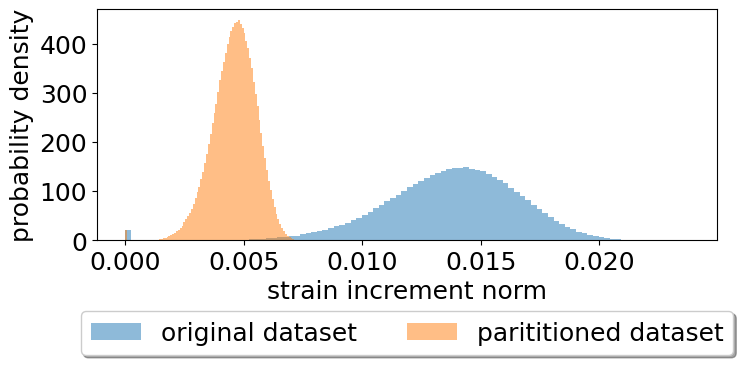}} 
    \caption{The estimated probability density function of strain components and strain increment norms for the original random-walk dataset and partitioned dataset.}
    \label{strain_pdf_new}
\end{figure}

The data generation method of He and Semnani \cite{he2023machine} is modified here to let small strain increments dominate the strain series. While $\mathbf Z$ converges with strain step refinement, as highlighted in Section \ref{sec:convergence}, the training stage is responsible for matching this converged state with the ground truth. In traditional numerical methods, the mathematical correspondence between the converged state and the ground truth naturally exists; however, a machine learning model learns such correspondence statistically from data. Therefore, a robust dataset for this problem must cover a sufficiently large range of total strain and consist of small strain increments. This motivates us to partition the dataset after the random-walk-based data generation is implemented \cite{he2023machine}. Partitioning the dataset into smaller increments (Algorithm \ref{dataset_partition}) does not change the total strain and strain path, but introduces smaller strain steps and increases the total number of steps (see Figures \ref{training_data_sample} and \ref{strain_pdf_new}). In this work, the training dataset for J2 plasticity has the following properties: Young's modulus $E = 50~\textrm{MPa}$, Poisson's ratio $\nu = 0.3$, uniaxial yield stress $\sigma_y = 1.2~\textrm{MPa}$, and linear plastic modulus $H' = 4$ MPa. In addition, isotropic ($\hat{\beta} = 1.0)$ and combined hardening ($\hat{\beta} = 0.5$) models are adopted, where $\hat{\beta}$ is the composite hardening parameter used for the combined isotropic and kinematic hardening model in J2 plasticity (see \cite{he2023machine} and \cite{borja2013plasticity}).
The Drucker-Prager material uses same values for $E$, $\nu$, and $\sigma_y$, with cohesion angle $\phi=30^{\circ}$ and dilation angle $\psi=25^{\circ}$. In particular, two cases are considered for training the Drucker-Prager INCDE model: in one case, INCDE outputs a regular 6-component stress, while the other case predicts pressure and deviatoric stress as output (i.e. 7 components in total). This aims to examine whether the volumetric-deviatoric decomposition can improve the performance of the surrogate model based on Drucker-Prager plasticity.

\begin{algorithm}
\caption{Uniform Partitioning of the Random-walk Dataset}
\label{dataset_partition}
  \begin{algorithmic}[1]
  \STATE \textbf{Input}: Set the original strain series as $\{\boldsymbol\epsilon^{<t>}\}$. This sample has $T$ steps and 6 components. Set the number of partitions, $C$.\\
  \STATE \textbf{Output}: $CT\times 6$ strain and stress series \\
    \FOR{$t=0,1,2,\cdots, T-1$}
      \FOR{$c=1,2,\cdots, C-1$}
        \STATE Use linear interpolation to determine $\boldsymbol{\epsilon}^{<t+\frac{c}{C}>} = \boldsymbol{\epsilon}^{<t>} + \frac{c}{C}\Delta\boldsymbol{\epsilon}^{<t+1>}$
      \ENDFOR
      \STATE Save values $\boldsymbol{\epsilon}^{<t>}$, $\boldsymbol{\epsilon}^{<t+\frac{1}{C}>}$, $\boldsymbol{\epsilon}^{<t+\frac{2}{C}>}$, ..., $\boldsymbol{\epsilon}^{<t+\frac{C-1}{C}>}$ to the new array for the refined strain series.
    \ENDFOR
    \STATE Determine the corresponding refined stress array and store it.
  \end{algorithmic}
\end{algorithm}

\subsection{Training}\label{sec:training}
Since components of inputs and outputs of the model differ in their magnitudes, strain and stress are normalized as follows to avoid biases in model parameters and improve training. The maximum absolute value of each strain or stress component is determined over all samples and time steps. Subsequently, the $k$th component in the $n$th sample at time step $t$ is normalized using the linear transformation below.
\begin{equation}\label{preprocessing}
    \begin{split}
        \bar\sigma_{n, k}^{<t>} = \frac{\sigma_{n, k}^{<t>}}{2\sigma_{k, \max}^{axial}}\ (k = 1, 2, 3) \quad , \quad
        \bar\sigma_{n, k}^{<t>} = \frac{\sigma_{n, k}^{<t>}}{2\sigma_{k, \max}^{shear}}\ (k = 4, 5, 6) \quad, \quad
        \bar\epsilon_{n, k}^{<t>} = \frac{\epsilon_{n, k}^{<t>}}{2\epsilon_{k, \max}}\quad , \quad
        \bar p_{n}^{<t>} = \frac{p_{n}^{<t>}}{2p_{\max}}
    \end{split}
\end{equation}
where
\begin{equation}\label{min_sigma}
    \begin{split}
        \sigma_{k, \max}^{axial} = \max_{n, t, k=1, 2, 3} |\sigma^{<t>}_{n, k}|\quad , \quad
        \sigma_{k, \max}^{shear} = \max_{n, t, k=4, 5, 6} |\sigma^{<t>}_{n, k}|\quad , \quad
        \epsilon_{k, \max} = \max_{n, t} |\epsilon^{<t>}_{n, k}|\quad , \quad
        p_{\max} = \max_{n, t} |p^{<t>}_{n}|
    \end{split}
\end{equation}
For Drucker-Prager plasticity, the same method for normalizing stress is applied to deviatoric stress. As a result, all strain and stress data are in $[-0.5, 0.5]$. During training and prediction, the transformed strain is given as input to the model. After prediction, the output is transformed back to obtain stress in the original space. We emphasize that this method does not shift the data distribution. A key advantage of scaling the data in range $[-0.5, 0.5]$ is that it prevents introducing bias in the initial map between zero strain and zero stress, since original zero values remain zeros after transformation.

\begin{figure}[ht]
    \centering
    \subfigure[]{\includegraphics[trim =0 0cm 0 0cm, clip, width=7.55cm]{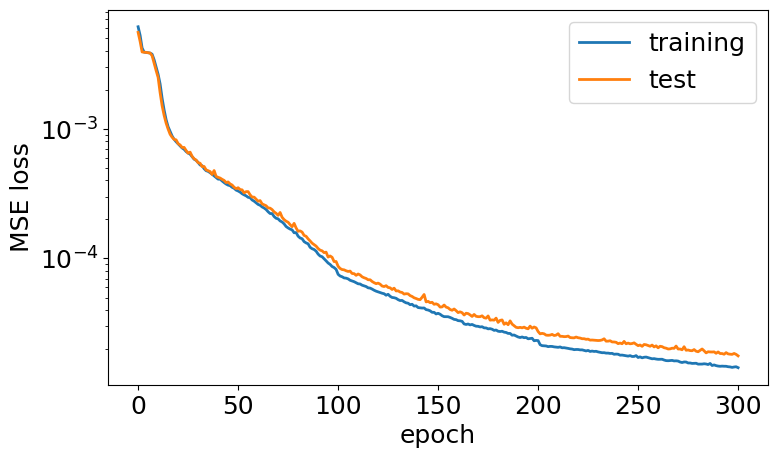} }
    \subfigure[]{\includegraphics[trim =0 0cm 0 0cm, clip, width=7.45cm]{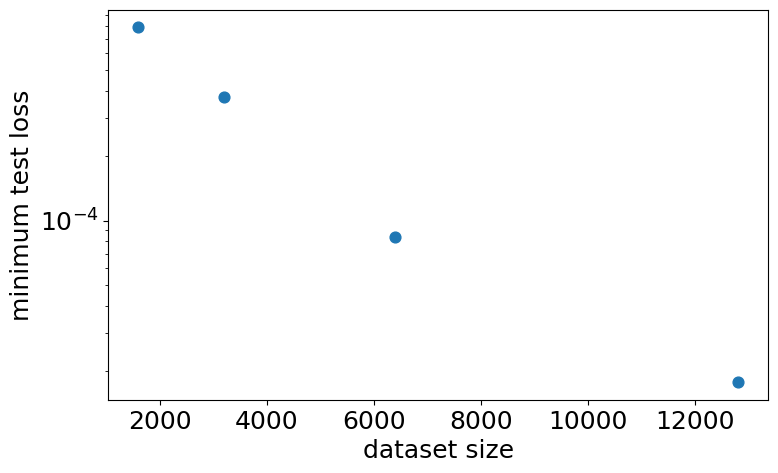}} 
    \caption{(a) Training and test losses using dataset size of 12,800 for the INCDE model trained with J2 plasticity (isotropic hardening); and (b) the minimum test loss versus size of the training dataset for the same model.}
    \label{training_test_loss}
\end{figure}

Mean-squared error (MSE) is chosen as the loss function for training as
\begin{equation}\label{MSE_loss}
    {\rm MSE} = \frac{1}{N} \sum_{n}^N\sum_{t}^T \Vert\bar{\boldsymbol\sigma}_{n}^{<t>} - \hat{\bar{\boldsymbol\sigma}}_{n}^{<t>}\Vert_2^2 \, .
\end{equation}
The training parameters are listed in Table \ref{training_params}. In addition, early stopping is implemented to prevent overfitting.
\begin{table}[ht]
    \centering
    \caption{Hyperparameters used for training.}
    \begin{tabular}{l l}
        \hline
        \textbf{Hyperparameter} & \textbf{Stress predictor}\\
        \hline
         Dataset size & J2: 12,800\\
          & DP: 18,000\\[5pt]
         Training/test ratio & 4:1\\[5pt]
         Epoch number & 300\\[5pt]
         Optimizer & Adam\\[5pt]
         Batch size & 128\\[5pt]
         $\Delta t$ & 0.2\\[5pt]
         Solver & midpoint\\[5pt]
          & $1\times10^{-3}$, Epoch 0 -- 100 \\
         Learning rate schedule & $5\times10^{-4}$, Epoch 100 -- 200\\
          & $2.5\times10^{-4}$, Epoch 200 -- 300\\
        \hline
    \end{tabular}
    \label{training_params}
\end{table}
Figure \ref{training_test_loss}(a) shows the training and test losses versus epochs for the INCDE model trained with isotropic J2 plasticity data. Figure \ref{training_test_loss}(b) illustrates the trend of the minimum test loss of the same INCDE model with respect to the dataset size. It is observed that the test loss decreases with more data. In this work, we select the dataset size of 12,800 since it provides a sufficiently small test loss. 

\section{Numerical Examples}\label{sec:examples}

\subsection{Stress Prediction at a Single Integration Point}
As explained in Section \ref{sec:convergence}, the hidden state and stress converge to nominal exact solutions upon refining $\Delta t$ and strain increments, whereas the error measured by the ground truth may not diminish continuously. This section provides two examples at a single integration point to verify the theories presented in Section \ref{sec:convergence}. The first example evaluates the INCDE model using a large set of test data to examine the effects of strain increment refinement on stress and hidden state errors and verify the stability of $\mathbf Z$ numerically. The second problem has a single tensile loading protocol, which investigates the effects of $\Delta t$, ODE solvers, and strain increment size on the stress and hidden state errors.

\subsubsection{Example 1: Effect of Strain Increment Size}\label{sec:example1}

\begin{figure} 
    \centering
    \subfigure[]{\raisebox{0.95cm}{\includegraphics[width=7.3cm]{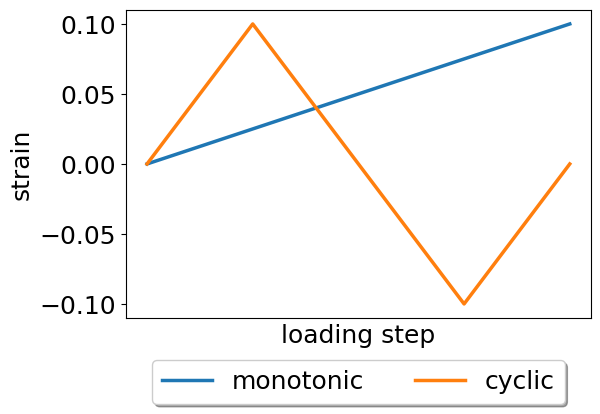}}}
    \subfigure[]{\includegraphics[width=7.1cm]{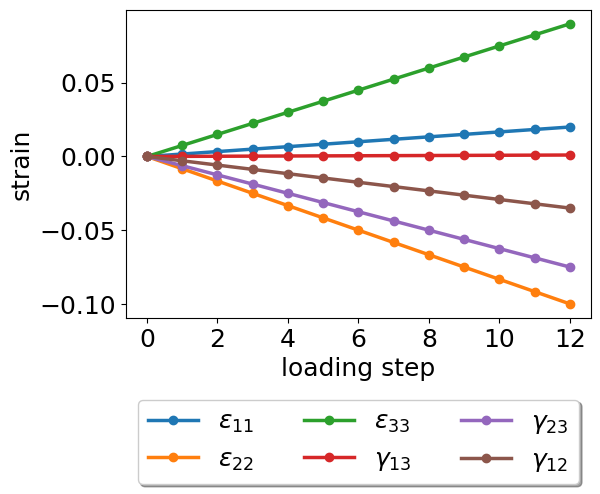}}
    \subfigure[]{\includegraphics[width=7.1cm]{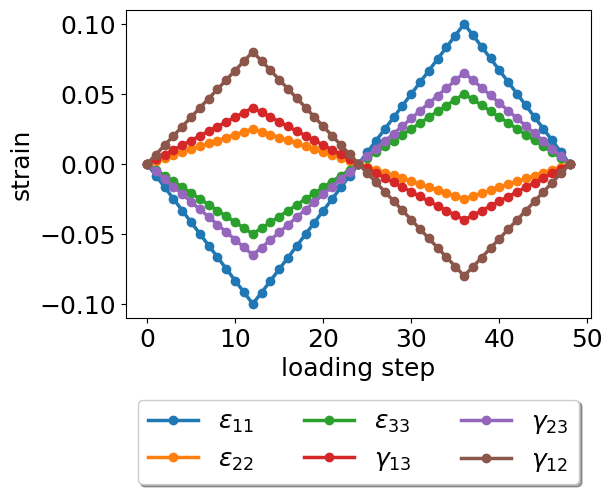}}
    \subfigure[]{\includegraphics[width=7.1cm]{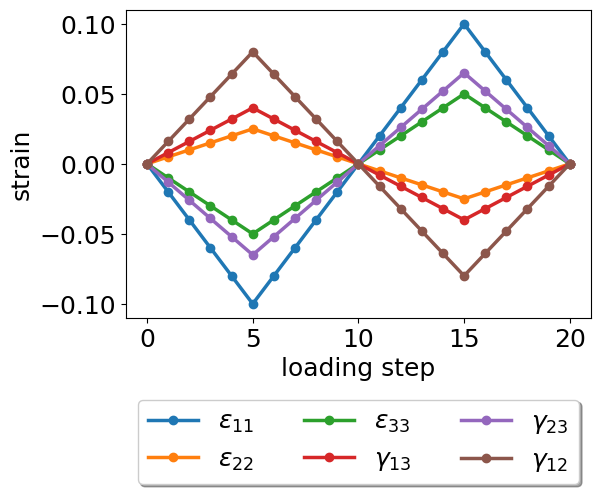}}
    \caption{Loading protocol plots: (a) standard monotonic and cyclic loading protocol shapes, (b) a sample of the monotonic strain set with $\Delta \epsilon_s = 0.008$, (c) a sample of the cyclic strain set with $\Delta \epsilon_s = 0.008$, and (d) a sample same as (c) with $\Delta \epsilon_s = 0.02$.}
    \label{fig:strain_protocol_plots}
\end{figure}

\begin{figure}
    \centering
    \includegraphics[width=15cm]{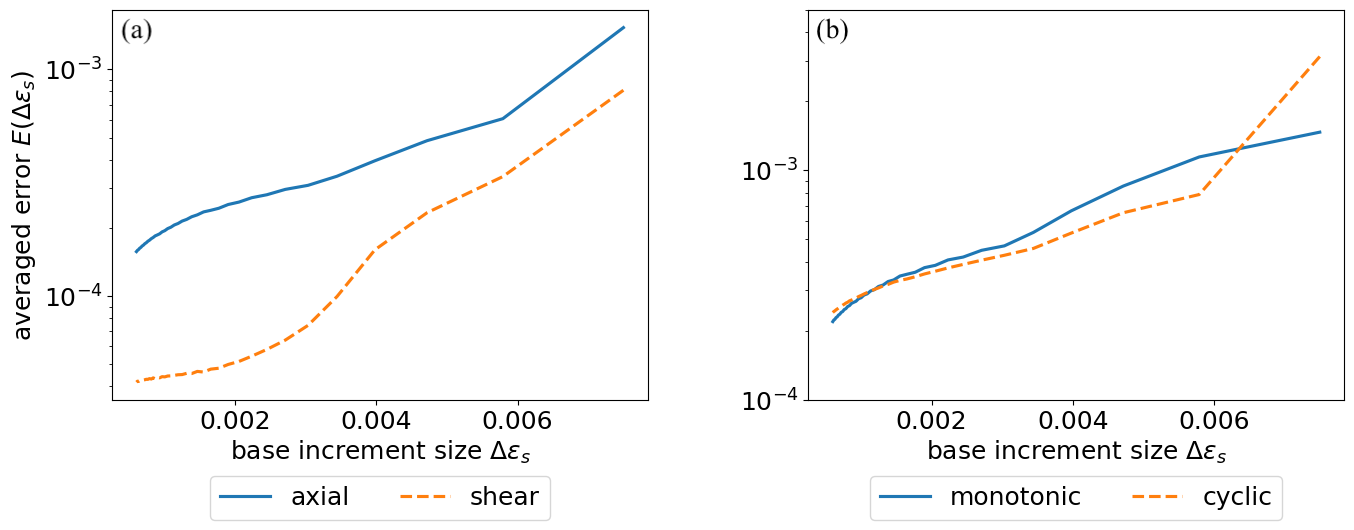}
    \caption{The average error $E(\Delta\epsilon_s)$ of stress predictions defined in Equation (\ref{eqn:sec611:err_metric}) versus base step size $\Delta\epsilon_s$ for (a) axial and shear components; and (b) monotonic and cyclic loading.}
    \label{fig:err_vs_step}
\end{figure}

This example studies the effects of strain
increment refinement on errors of the predicted stresses compared to the ground truth (return mapping). For this purpose, an experiment is designed as follows.
\begin{enumerate}
    \item First, two strain datasets are generated for monotonic and cyclic loading protocols. Each set has $N=1128$ protocols consisting of 6 strain components. The shapes of strain components in the monotonic and cyclic protocols are schematically displayed in Figure \ref{fig:strain_protocol_plots}a. In each set, the paths of strain components differ in their initial loading directions and peak strain $\epsilon_{max}$. For each component of a given sample, $\epsilon_{max}$ is drawn from a uniform distribution $U(0, 0.1)$ and the loading direction is randomly chosen as $-1$ or $1$. Examples of strain protocols are shown in Figures \ref{fig:strain_protocol_plots}b, \ref{fig:strain_protocol_plots}c, and \ref{fig:strain_protocol_plots}d.
    \item Second, 40 values of the base increment size, $\Delta \epsilon_s$, are selected ranging from $6\times 10^{-4}$ to $7\times 10^{-3}$ to discretize the generated strain protocols. When using $\Delta \epsilon_s$ to discretize a protocol with peak strain of $\epsilon_{max}$, the discretization step size is scaled to $\Delta \epsilon = \Delta \epsilon_s \epsilon_{max} / 0.1$. This adjustment allows different components of the discretized series to have the same lengths (i.e. number of time steps) equal to
    \begin{equation}
        T(\Delta\epsilon_s) = \begin{cases}
            \lfloor 0.1/\Delta \epsilon_s \rfloor \quad (\textrm{monotonic set})\\
            4 \lfloor 0.1/\Delta \epsilon_s\rfloor \quad (\textrm{cyclic set}) \, ,
        \end{cases}
    \end{equation}
    where $\lfloor \cdot \rfloor$ denotes the floor function. Figures \ref{fig:strain_protocol_plots}c and \ref{fig:strain_protocol_plots}d show same loading protocols but discretized using different values of $\Delta\epsilon_s$.
    \item Finally, for each $\Delta \epsilon_s$, the $N$ protocols are discretized and input to the trained INCDE model. Subsequently, a $N\times T\times 6$ stress tensor $\hat{\boldsymbol\sigma}$ is obtained for each $\Delta\epsilon_s$ and compared with the corresponding return mapping solution $\boldsymbol\sigma_{\textrm{ref}}$. We define an error metric as follows to assess the error of stress prediction for every $\Delta\epsilon_s$:
    \begin{equation}\label{eqn:sec611:err_metric}
        E(\Delta \epsilon_s) = \frac{\Vert \hat{\boldsymbol\sigma}(\Delta\epsilon_s) - \boldsymbol{\sigma}_{\textrm{ref}} \Vert}{NT(\Delta\epsilon_s)} \, .
    \end{equation}
\end{enumerate} 
Figure \ref{fig:err_vs_step}a shows $E(\Delta \epsilon_s)$ for axial/shear stress components using both monotonic and cyclic datasets, while Figure \ref{fig:err_vs_step}b shows the errors for monotonic and cyclic loading separately. The test set used here is considered difficult for the model since the loading patterns were not included in the training set. Nevertheless, it is observed that the model performs well and produces small errors. Figure \ref{fig:err_vs_step}b shows that the range of errors in predictions under cyclic loading is larger than monotonic loading, since the strain patterns are more complex under cyclic load. Moreover, the error $E(\Delta \epsilon_s)$ decreases with refined discretization represented by $\Delta\epsilon_s$, which shows the self-consistency and convergence properties of the method as discussed in Sections \ref{sec:convergence} and \ref{sec:incde-advantage}.

\begin{figure}
    \centering
    \subfigure[]{\includegraphics[width=7.4cm]{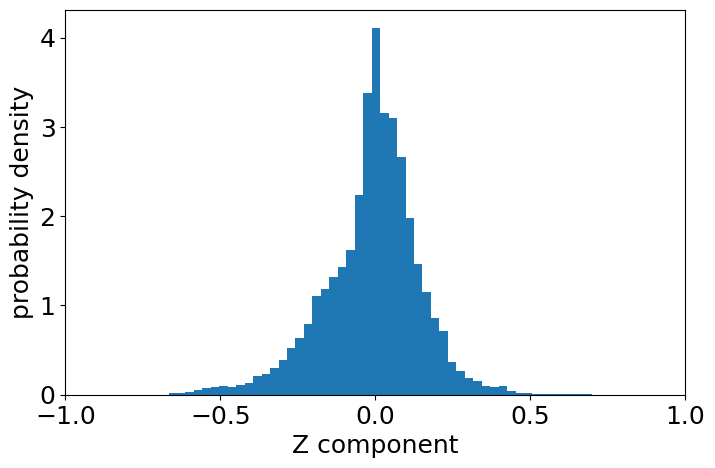}} 
    \subfigure[]{\includegraphics[width=7.6cm]{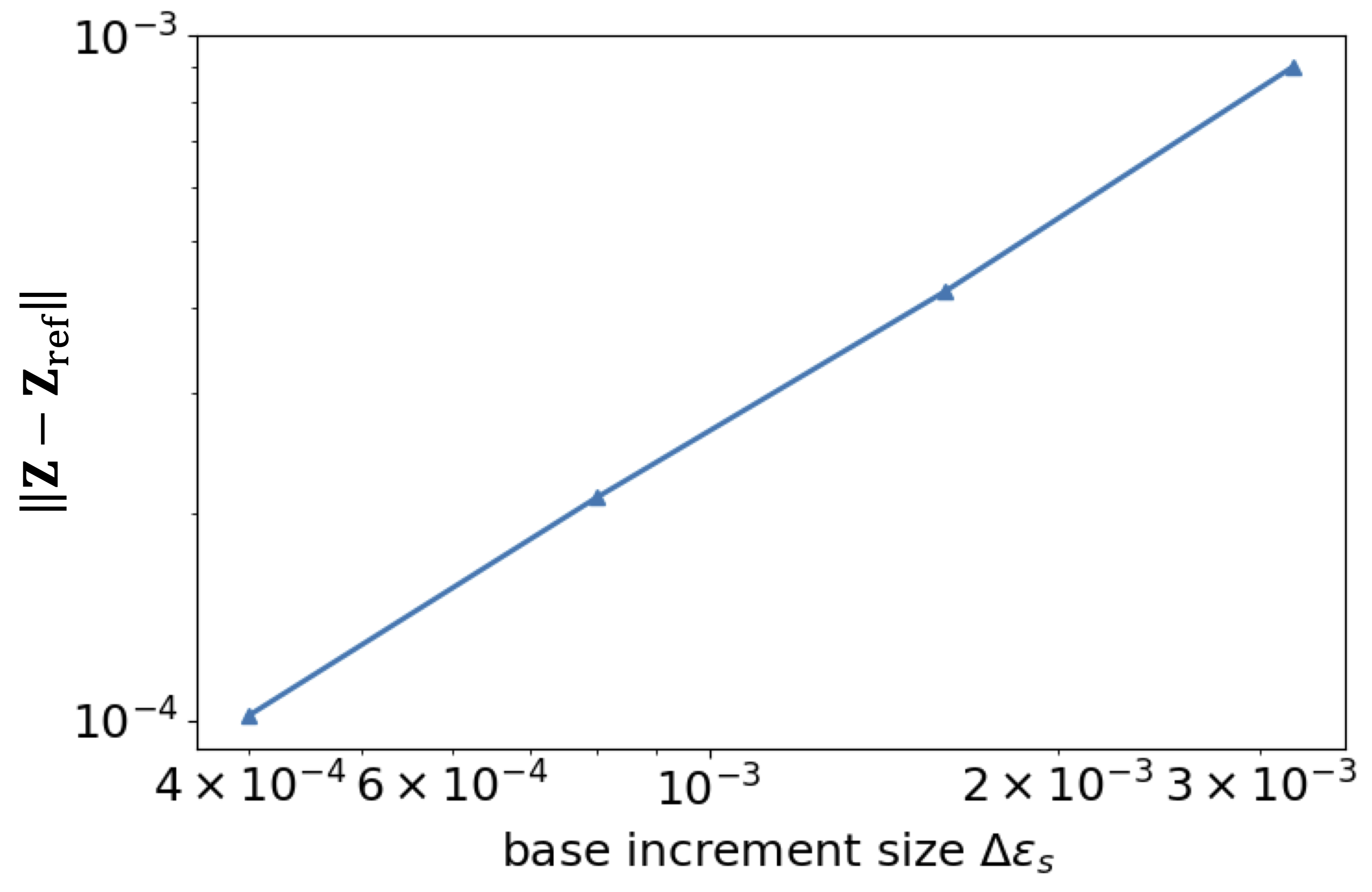} }
    \caption{(a) Estimated probability density function of components of $\mathbf Z$ in loading protocol tests of Example 1; and (b) error $\Vert \mathbf Z - \mathbf Z_{\textrm{ref}} \Vert$ versus base strain increment size showing the convergence trajectory of $\mathbf Z$.}
    \label{num_convergence_stability}
\end{figure}

Stability and convergence of the hidden state of the INCDE model are also studied using the test dataset described above. We demonstrate stability by obtaining the estimated probability density plots of the components of $\mathbf Z$ during testing (Figure \ref{num_convergence_stability}a). It is observed that the distribution of $\mathbf Z$ is in the range $(-1, 1)$, which agrees with the theoretical analysis of Section \ref{sec:stabilize-incde}.
To study convergence, we first define an error metric for $\mathbf Z$. For this purpose, 300 strain loading protocols are chosen from the validation dataset. For each loading protocol, different increment sizes, $\Delta\epsilon_s = \{2\times10^{-4}, 4\times10^{-4}, 8\times10^{-4}, 1.6\times10^{-3}, 3.2\times10^{-3}\}$, are selected to uniformly discretize the strain series. $\mathbf Z$ series computed with the smallest discretization (i.e., $\Delta\epsilon_s = 2\times10^{-4}$) is defined as the reference trajectory $\mathbf Z_{\textrm{ref}}$ for each loading protocol. Subsequently, the results computed with the remaining increment sizes are determined and their errors are computed as $\Vert \mathbf Z - \mathbf Z_{\textrm{ref}} \Vert$. The errors calculated for each increment size are averaged and plotted versus increment size (Figure \ref{num_convergence_stability}b). It is observed that the average error drops consistently with decreasing strain increment sizes, demonstrating convergence of the hidden state of the INCDE model.

\subsubsection{Example 2: Effects of Time Step Size and ODE Solvers}\label{sec:example2}
\begin{figure}[ht]
    \centering
    \includegraphics[width=15cm]{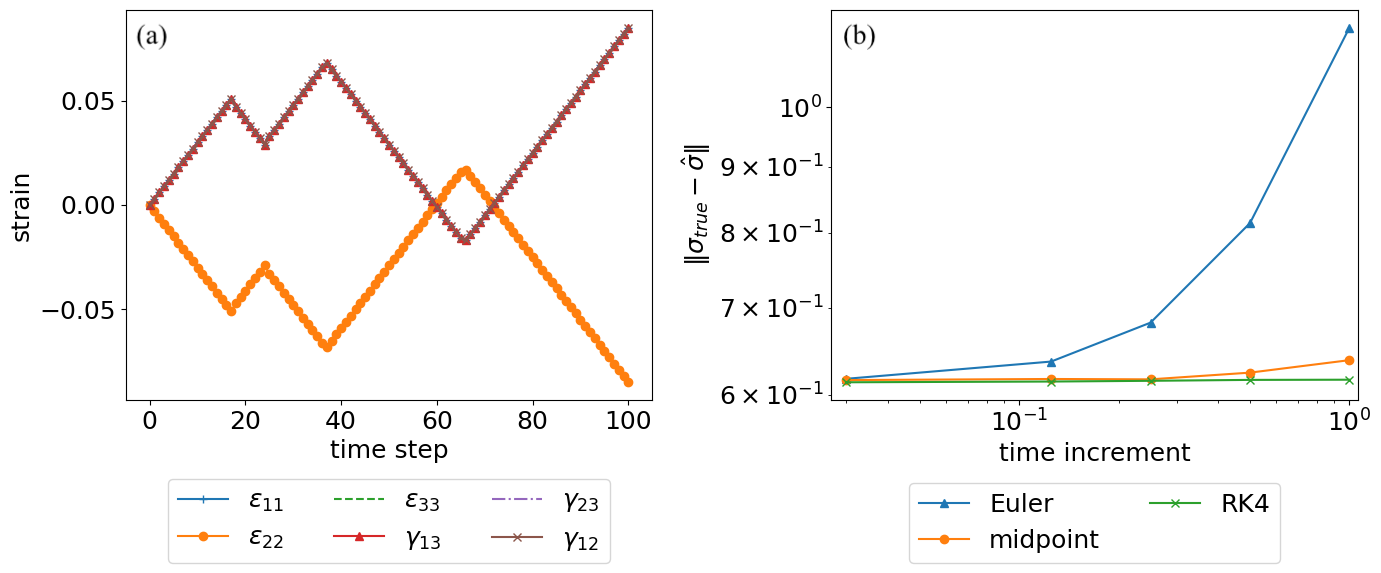}
    \caption{(a) Loading protocol for the stress prediction problem in Example 2; and (b) convergence of the INCDE predicted stress compared with the true stress versus time increment size for forward Euler, RK4 and midpoint methods.}
    \label{gpt_example2_loading_protocol_and_rm_stress}
\end{figure}

In this example, the strain loading protocol shown in Figure \ref{gpt_example2_loading_protocol_and_rm_stress}a is applied to an integration point. All components of the strain series are identical except $\epsilon_{22}$ which has a reversed direction. This problem aims to study the effects of time increment $\Delta t$ and ODE solvers; therefore, absolute values of all components of $\Delta\boldsymbol\epsilon$ are fixed to 0.003. Time increments are 
$\Delta t = 0.03, 0.125, 0.25, 0.5, 1.0$ and ODE solvers are selected as the forward Euler, RK4, and the midpoints methods. 

Figure \ref{gpt_example2_loading_protocol_and_rm_stress}b shows error of the predicted stress compared to the stress computed using conventional plasticity, which is the ground truth. It can be seen that stress predictions converge with refining the time steps. However, the convergence rate is not the same as the theoretical convergence rate of numerical ODE solvers. This is due to the fact that during training, the converged $\mathbf Z$ is used to predict stress, which is then compared with the true stress to calculate the MSE loss. During training, the midpoint method with $\Delta t = 0.2$ is used as the ODE solver, which means that $\mathbf Z$ solved using a method with an order of accuracy at $\text{O}(0.2^2) = \text{O}(0.04)$ is treated as a sufficiently converged solution and mapped to the ground truth. Therefore, during testing, a numerical method with a higher order of accuracy than $\text{O}(0.04)$ may not reduce stress errors compared to the ground truth any further. Accordingly, as observed in Figure \ref{gpt_example2_loading_protocol_and_rm_stress}b, small time increments of RK4 solver do not reduce the errors considerably. The Euler solver continues to improve with refining time increments until it reaches a plateau at $\Delta t = 0.03$ with almost the same accuracy as the midpoint method and RK4. 

\begin{figure}[ht]
    \centering
    \includegraphics[width=15cm]{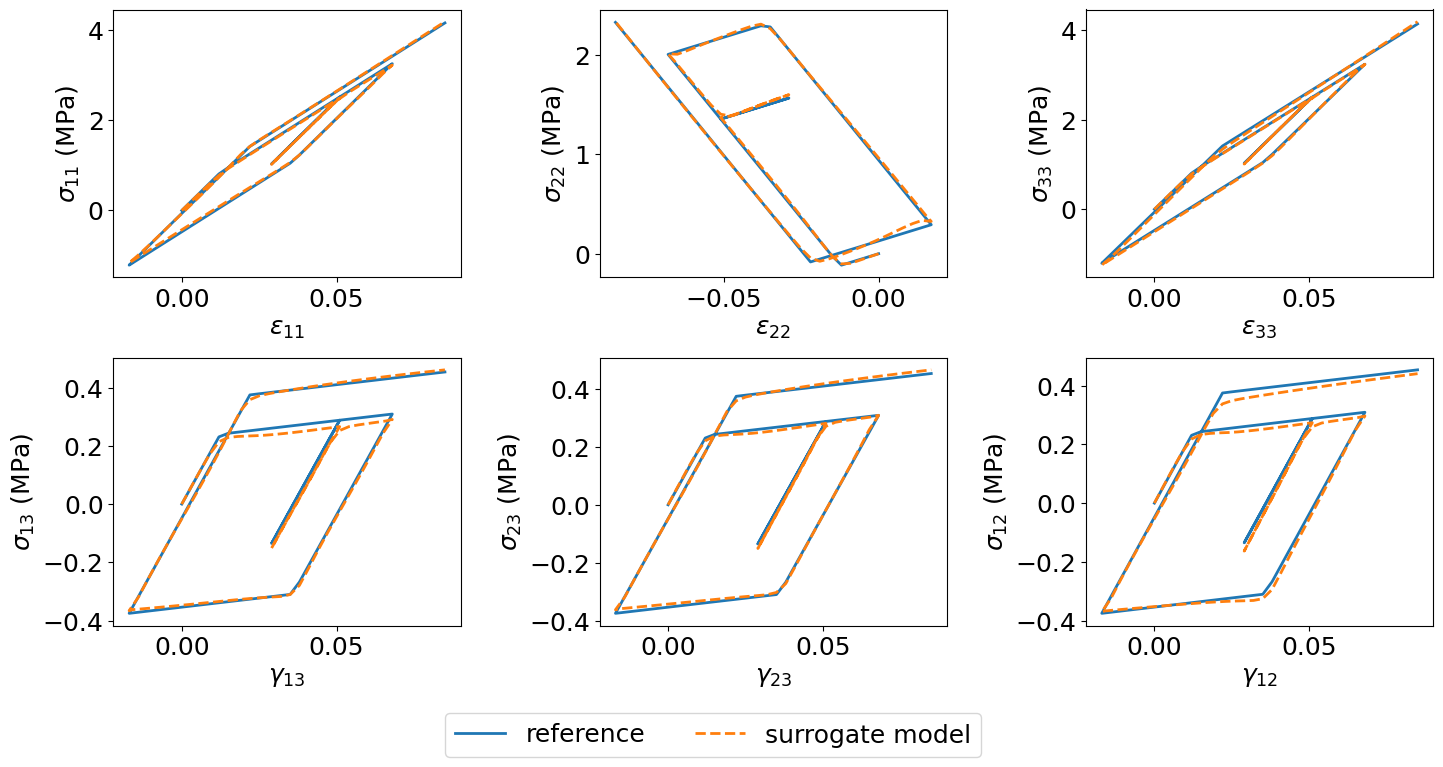}
    \caption{Stress-strain curves obtained using conventional J2 plasticity with isotropic hardening (return mapping algorithm) and the INCDE model using RK4 with $\Delta t = 0.1$.}
    \label{gpt_example2_ref_stress_strain}
\end{figure}

\begin{figure}[ht]
    \centering
    \includegraphics[width=15cm]{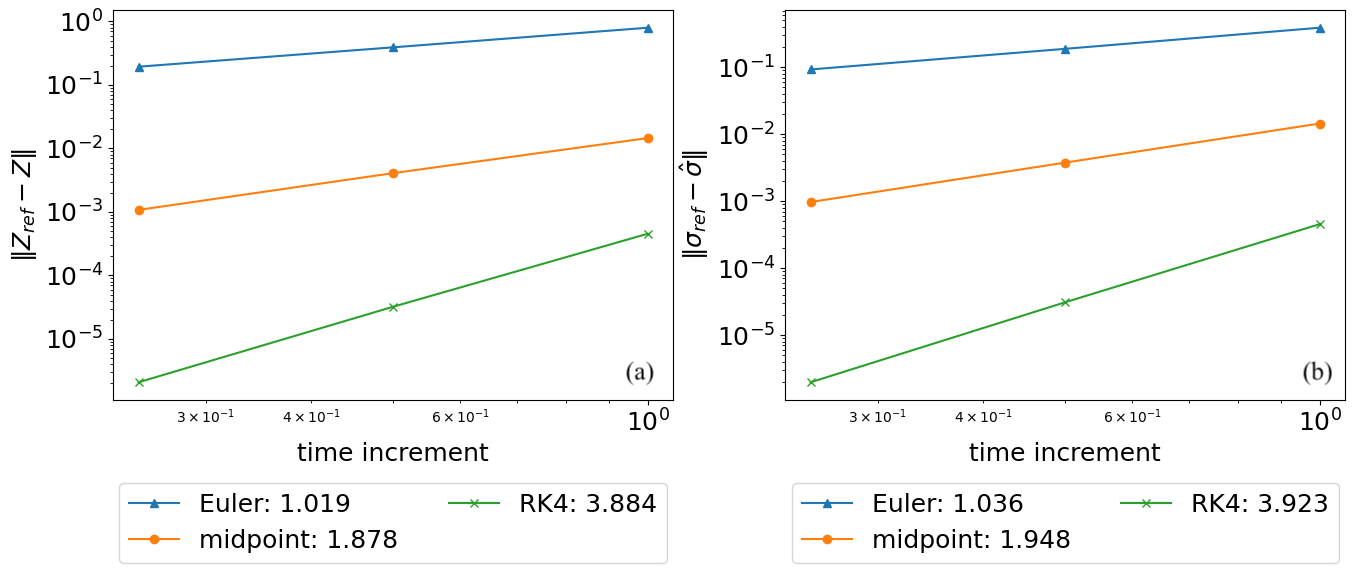}
    \caption{Convergence rate of (a) hidden state $\mathbf Z$ compared with $\mathbf Z_{\textrm{ref}}$; and (b) predicted stress compared with $\boldsymbol{\sigma}_{\textrm{ref}}$ with respect to time increment size for forward Euler, midpoint and RK4 methods. The slopes of the logarithmic lines are appended to each label.}
    \label{gpt_example2_numerical_convergence}
\end{figure}

\begin{figure}[ht]
    \centering
    \includegraphics[width=8cm]{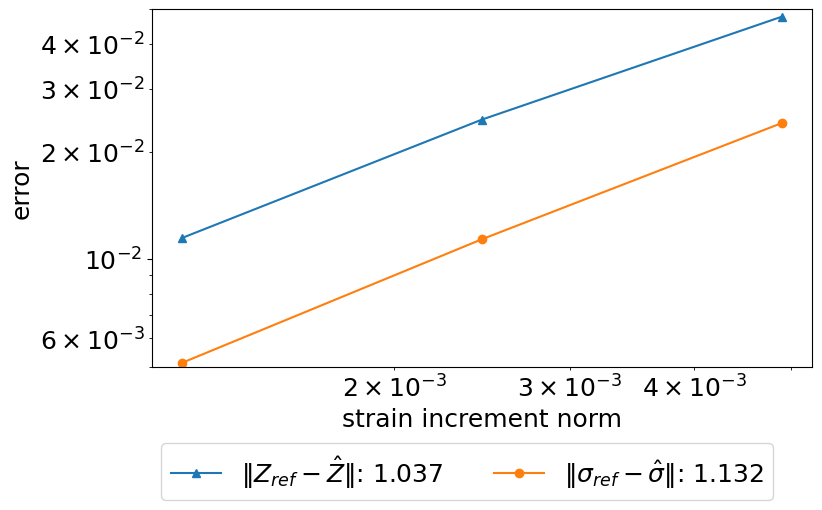}
    \caption{Error of the hidden state and predicted stress versus the norm of uniform strain increments. RK4 method with $\Delta t = 0.1$ is used during this test.}
    \label{gpt_example2_numerical_convergence_versus_dep}
\end{figure}

Next, we treat the hidden state and stress obtained using RK4 with $\Delta t = 0.1$ (Figure \ref{gpt_example2_ref_stress_strain}) as the nominally exact solutions ($\mathbf Z_{\textrm{ref}}$ and $\boldsymbol{\sigma}_{\textrm{ref}}$) to obtain the prediction errors and convergence plots. Figure \ref{gpt_example2_numerical_convergence} shows that $\mathbf Z$ and $\hat{\boldsymbol{\sigma}}$ converge in accordance with the theoretical convergence rates of numerical ODE solvers. Finally, we use the RK4 solver with $\Delta t = 0.1$ to study convergence of  $\mathbf Z$ and $\hat{\boldsymbol{\sigma}}$ with respect to strain increment size. Errors are calculated via comparison with the nominal solution corresponding to $\Vert\Delta\boldsymbol\epsilon\Vert = 1.225 \times 10^{-4}$. A linear convergence rate is observed in Figure \ref{gpt_example2_numerical_convergence_versus_dep} for both $\mathbf Z$ and $\hat{\boldsymbol{\sigma}}$, which agrees with the theoretical analysis of Section \ref{sec:convergence}.

This example problem verifies two main characteristics of the INCDE stress predictor. On one hand, the INCDE model preserves numerical convergence properties of ODEs. On the other hand, the link between hidden state $\mathbf Z$ and true stress is learned statistically from data; therefore, the theoretical error analysis is not applicable to the error calculated based on ground truth. 


\subsubsection{Example 3: Integration Point simulation Using J2 Plasticity with Combined Hardening}

\begin{figure}[ht]
    \centering
    \includegraphics[width=15cm]{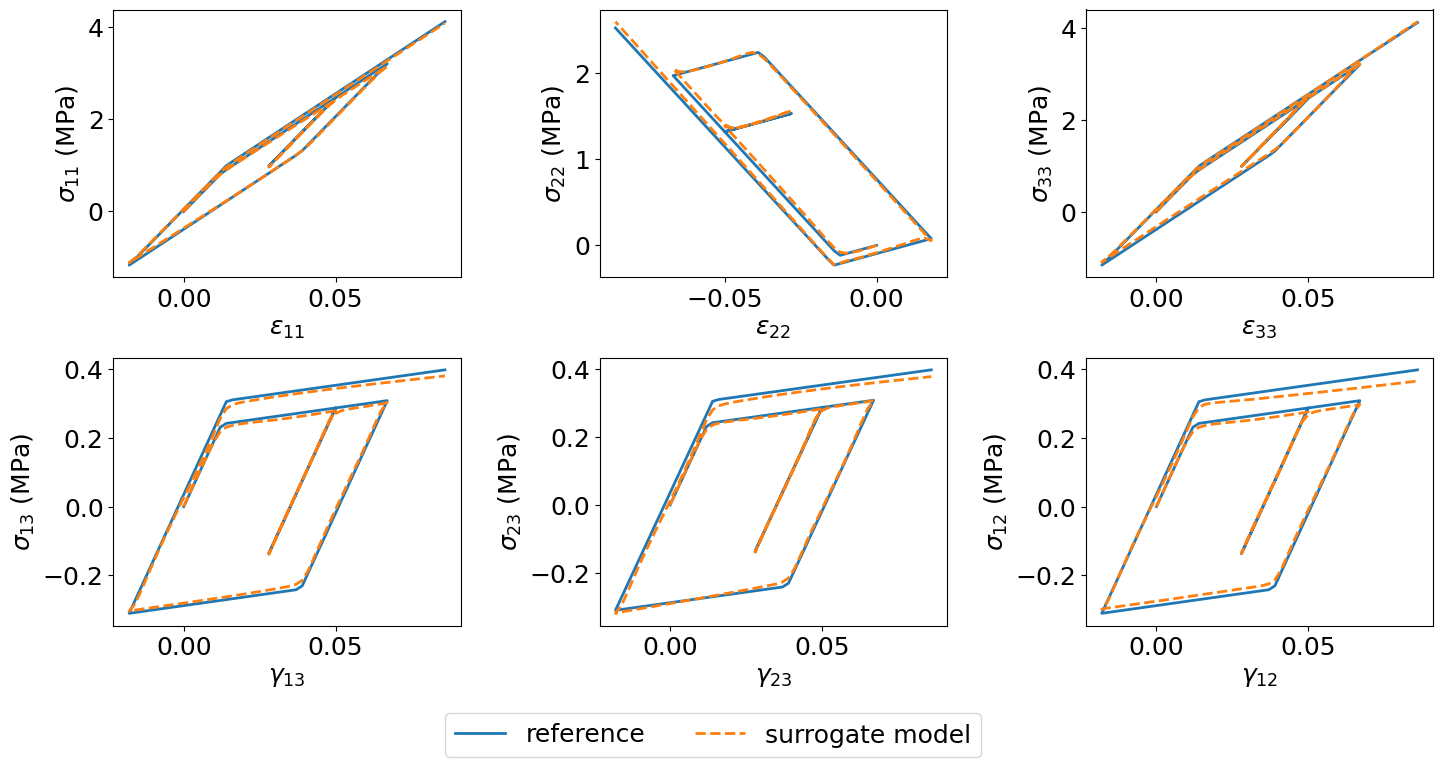}
    \caption{Stress-strain curves obtained using combined hardening J2 plasticity ($\hat{\beta} = 0.5$) and the corresponding INCDE model using RK4 with $\Delta t = 0.1$.}
    \label{gpt_beta05_stress_strain}
\end{figure}

This example aims to test the performance of the INCDE model trained for J2 plasticity with combined hardening ($\hat{\beta}=0.5$). The strain loading protocol shown in Figure \ref{gpt_example2_loading_protocol_and_rm_stress}a is applied to an integration point, and the problem is simulated with the RK4 solver using $\Delta t = 0.1$. Results are shown in Figure \ref{gpt_beta05_stress_strain}, and it is observed that the INCDE model captures the path-dependent material behavior and achieves high accuracy in loading/unloading with back stress.

\subsection{FE Simulations for Boundary Value Problems}\label{sec:FE-example}
In this section, we aim to investigate the performance of the INCDE-based models in simulating boundary value problems. For this purpose, the INCDE stress predictor replaces the conventional J2/Drucker-Prager plasticity return mapping algorithm within a nonlinear finite element solver. Subsequently, two boundary value problems are solved with the INCDE-based finite element method.

The nonlinear system derived from the finite element discretization is solved using Broyden's method, which is a three-dimensional extension of the secant method \cite{Broyden:1965vv}. For each loading increment, only the global tangential stiffness matrix $\mathbf K_t$ in the first iteration is assembled with $\hat{\mathbf C}_t$ from backpropagation (see Figure \ref{INCDE_backprop}). In the subsequent iterations, $\mathbf K_t$ is updated using Broyden's secant matrix. Broyden's method converges superlinearly but bypasses the most expensive backpropagation steps, so it is overall faster than Newton's method. We adopt the following convergence criterion for the unbalanced force $\mathbf r_f$ at free degrees of freedom 
\begin{equation}\label{convergence_criterion}
    \Vert \mathbf r_f \Vert_2 \leq \tau_0\ \textrm{and}\ \Vert \mathbf r_f \Vert_2 \leq \tau_r r_0 \, ,
\end{equation}
where $r_0$ is the norm of $\mathbf r_f$ in the first iteration of each increment. The tolerances $\tau_0=1\times10^{-5}$ and $\tau_r=1\times10^{-5}$ control the absolute and relative magnitudes of the unbalanced force, respectively. The details of the FE implementation is summarized in Algorithm \ref{FE_surrogate_model_implementation}. Evaluation of the residual force $\mathbf r_f$ in this algorithm is discussed in \cite{he2023machine}.

\begin{algorithm}
\caption{Finite Element Implementation Using Surrogate Models and Broyden's Method}
\label{FE_surrogate_model_implementation}
  \begin{algorithmic}[1]
  \STATE {Initialize $\mathbf u_0 = \mathbf 0$.}
    \FOR{oading step $n=1,2,\cdots$}
      \STATE Initialize $\Delta \mathbf u^{<0>} = \mathbf 0$.
      \FOR{iteration $i= 1,2,\cdots$}
        \STATE Evaluate the residual force $\mathbf r_f^{<i>} = \mathbf r_f(\mathbf u_{n} + \Delta\mathbf u^{<i>})$ at free DOFs.
        \IF{$\Vert \mathbf r_f \Vert_2 \leq \tau_0\ \textrm{and}\ \Vert \mathbf r_f \Vert_2 \leq \tau_r r_0$}
            \STATE Update $\mathbf u_{n+1} = \mathbf u_{n} + \Delta \mathbf u^{<i>}$
            \STATE Update material state variables at integration points.   // \textit{Hidden state $\mathbf Z$ is updated.}
            \STATE \textbf{break}
        \ENDIF
        \IF{i == 0}
            \STATE Compute material tangential stiffness $\hat{\mathbf C}_t$ at integration points. // \textit{Backpropagation is performed.}
            \STATE Assemble the exact global tangential stiffness $\mathbf K_t^{<0>}$.  
            \STATE Initialize secant stiffness $\mathbf B^{<0>} = \mathbf K_t^{<0>}$
        \ELSE
            \STATE Compute $ D\mathbf u = \Delta \mathbf u^{<i>} - \Delta \mathbf u^{<i - 1>}$.
            \STATE Update $(\mathbf B^{<i>})^{-1} = (\mathbf B^{<i-1>})^{-1} + \frac{D \mathbf u - (\mathbf B^{<i-1>})^{-1}\Delta \mathbf r_f^{<i>}}{D \mathbf u^T(\mathbf B^{<i-1>})^{-1}\Delta \mathbf r_f^{<i>}}D \mathbf u^T(\mathbf B^{<i-1>})^{-1}$.  // \textit{Sherman–Morrison formula}
        \ENDIF
        \STATE Solve $\mathbf B^{<i>}\delta \Delta \mathbf u^{<i>} = \mathbf r_f^{<i>}$ for $\delta \Delta \mathbf u^{<i>}$.
        \STATE $\Delta\mathbf u^{<i+1>} = \Delta\mathbf u^{<i>} - \delta \Delta \mathbf u^{<i>}$.
      \ENDFOR
    \ENDFOR
  \end{algorithmic}

\end{algorithm}

\subsubsection{Coupon with Imperfection}\label{sec:coupon}
\begin{figure}[ht]
    \centering
    \includegraphics[width=0.8\textwidth]{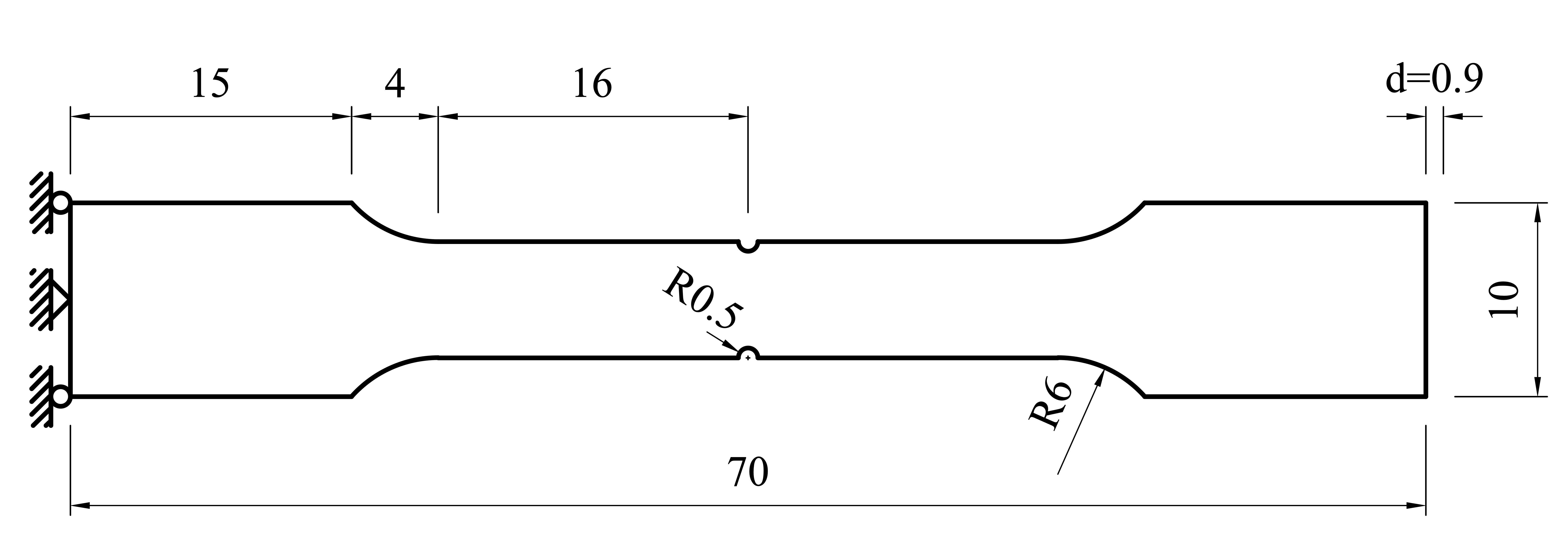}
    \caption{Geometry and boundary conditions of the coupon problem with imperfection.}
    \label{coupon_dimension}
\end{figure}

\begin{figure}[ht]
    \centering
    \subfigure[]{\includegraphics[width=6.75cm]{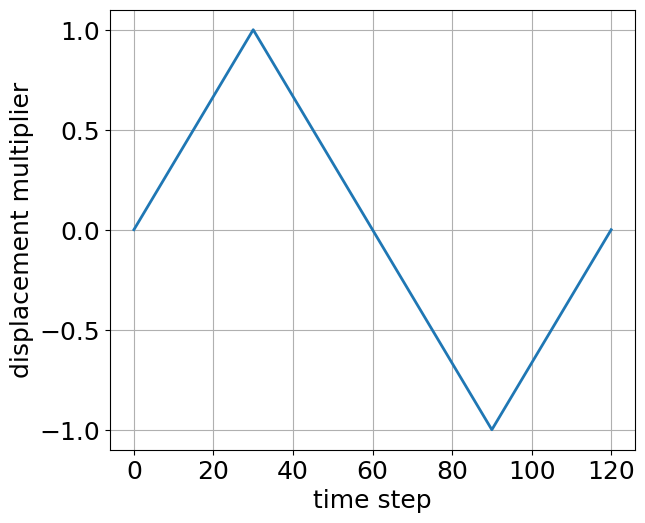}} 
    \subfigure[]{\includegraphics[width=8.25cm]{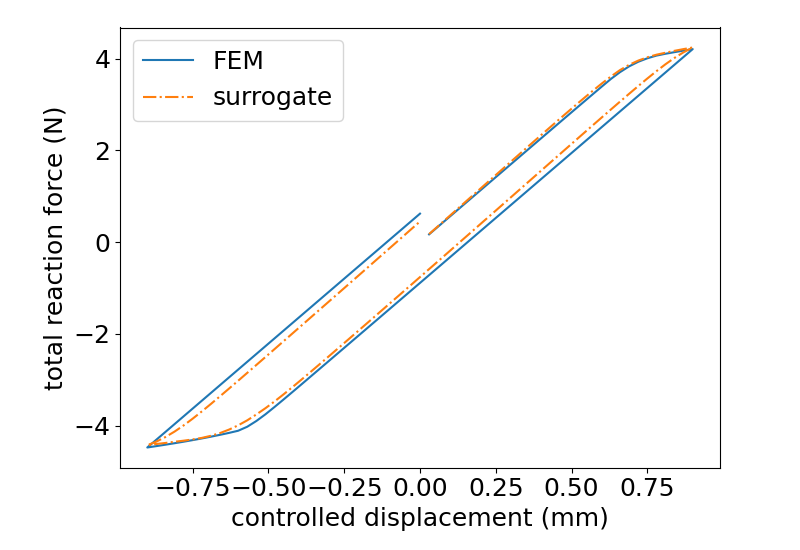}}
    \caption{(a) The loading protocol applied to the coupon problem, and (b) the total reaction force versus controlled displacement.}
    \label{coupon_loading}
\end{figure}

A thick coupon specimen is shown in Figure \ref{coupon_dimension}. There are two symmetric circular imperfections in the middle of the coupon. A uniform displacement $d=0.9~\textrm{mm}$ is applied to the right side. Figure \ref{coupon_loading}a shows the time evolution of the multiplier used to scale $d$. From Step 0 to Step 25, the coupon is stretched. When fully loaded, the coupon is unloaded, compressed in the negative direction, then unloaded to the original position. Due to the symmetry of the problem, only one quarter of the coupon is modeled. Since the coupon is thick, plane strain 4-node quadrilateral element is used to mesh the coupon. The material behavior is J2 plasticity with isotropic hardening, Young's modulus $E = 50~\textrm{MPa}$, Poisson's ratio $\nu = 0.3$, initial yield stress $\sigma_y = 1.2~\textrm{MPa}$, and linear plastic modulus $H' = 4~\textrm{MPa}$. The INCDE model uses the midpoint method with $\Delta t =0.5$.

\begin{figure}[ht]
    \centering
    \subfigure[]{\includegraphics[trim =1.8cm 0cm 1.8cm 0.5cm, clip, width=7.5cm]{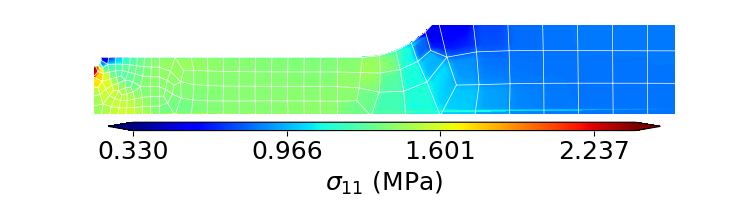}} 
    \subfigure[]{\includegraphics[trim =1.8cm 0cm 1.8cm 0.5cm, clip, width=7.5cm]{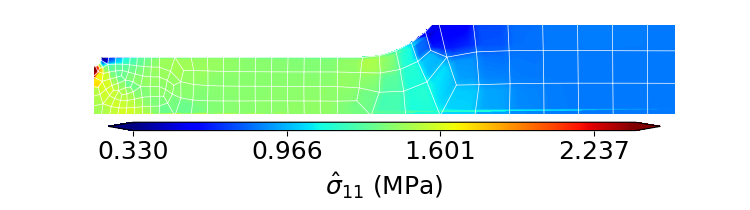}}
    \subfigure[]{\includegraphics[trim =1.8cm 0cm 1.8cm 0.5cm, clip, width=7.5cm]{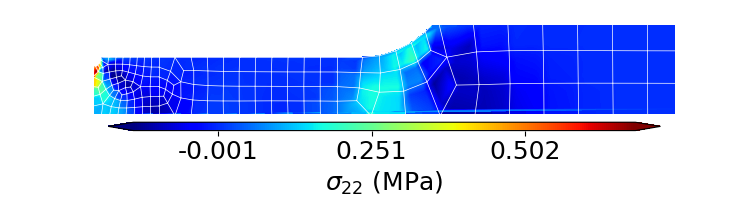}} 
    \subfigure[]{\includegraphics[trim =1.8cm 0cm 1.8cm 0.5cm, clip, width=7.5cm]{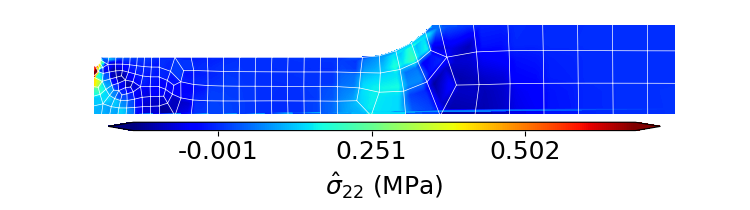}}
    \subfigure[]{\includegraphics[trim =1.8cm 0cm 1.8cm 0.5cm, clip, width=7.5cm]{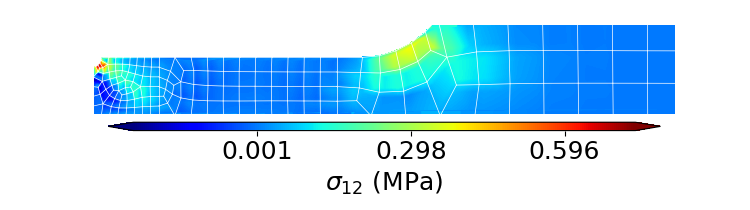}} 
    \subfigure[]{\includegraphics[trim =1.8cm 0cm 1.8cm 0.5cm, clip, width=7.5cm]{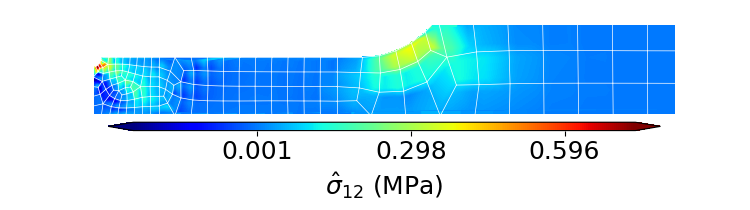}}
    \caption{Contour plots of the stress field at the maximum positive displacement. The left column displays the results from standard J2 plasticity, whereas the right column shows results from the surrogate model.}
    \label{coupon_stress_contour}
\end{figure}

Figure \ref{coupon_loading}b shows the reaction force versus controlled displacement at the right edge of the coupon obtained using the INCDE model compared with the standard FE solution. It is observed that the surrogate model captures the global elasto-plastic behavior very well. Figure \ref{coupon_stress_contour} visualizes the contour plots of stress components at the state of maximum stretch $d=0.9~\textrm{mm}$. A very good match is observed between the INCDE and the standard FE results. Particularly, the stress concentrations are predicted well at the imperfection and fillet. In addition, the INCDE model captures the distribution of $\sigma_{11}$ in the neck and the shear band in the vicinity of the imperfection. These observations indicate that the numerical solution of the INCDE reaches the same state of plasticity when the structure is fully loaded.


\subsubsection{Square Plate with a Hole}\label{sec:square-plate}
\begin{figure}[ht]
    \centering
    \subfigure[]{\includegraphics[width=7.5cm]{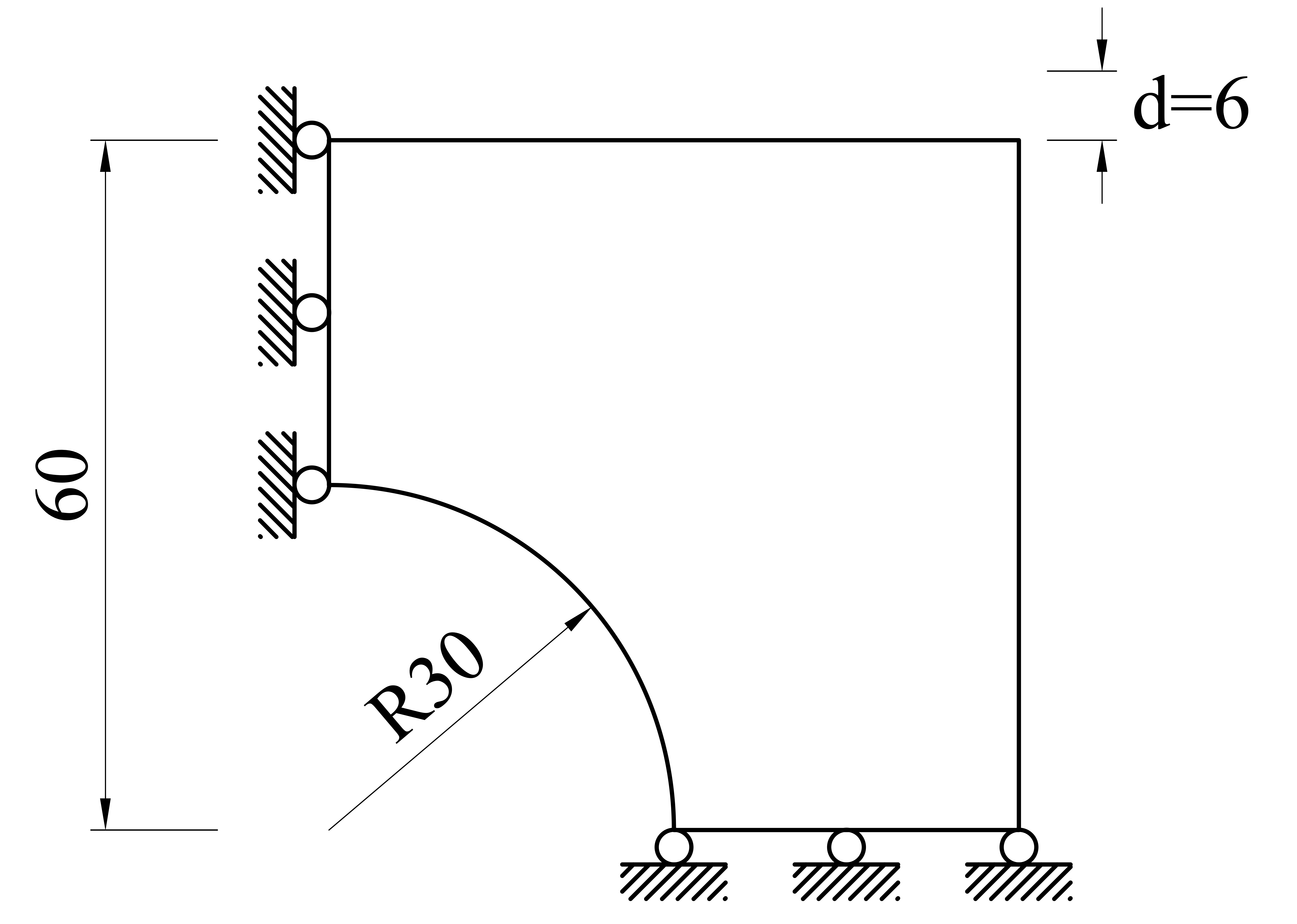}} 
    \subfigure[]{\includegraphics[width=4.6cm]{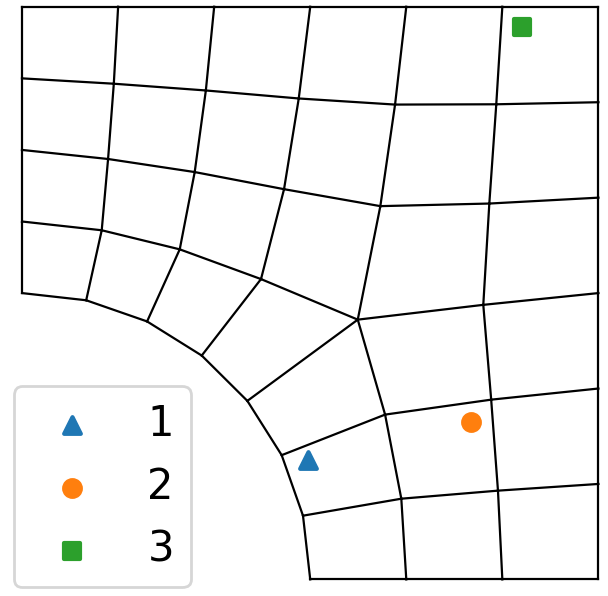}} 
    \caption{(a) Geometry and boundary conditions of the square plate problem; and (b) mesh used for the problem with the marked integration points used to obtain stress-strain curves.}
    \label{quarter_plate_bc}
\end{figure}

\begin{figure}
    \centering
    \subfigure[]{\includegraphics[trim =0 0cm 0 0cm, clip, width=6.75cm]{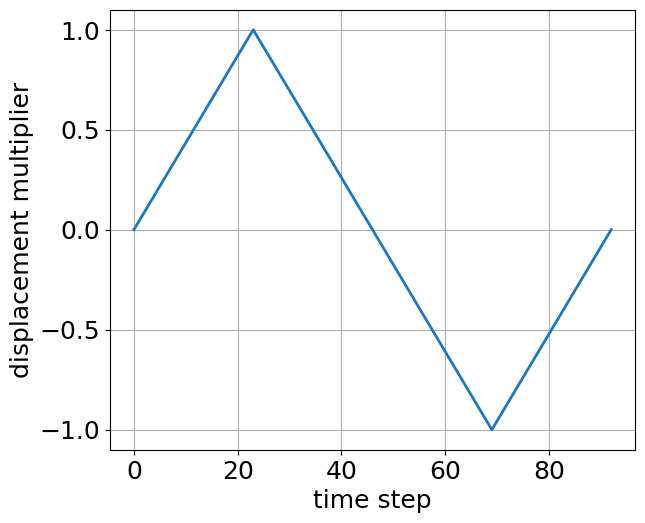}} 
    \subfigure[]{\includegraphics[trim =0 0cm 0 0cm, clip, width=7.1cm]{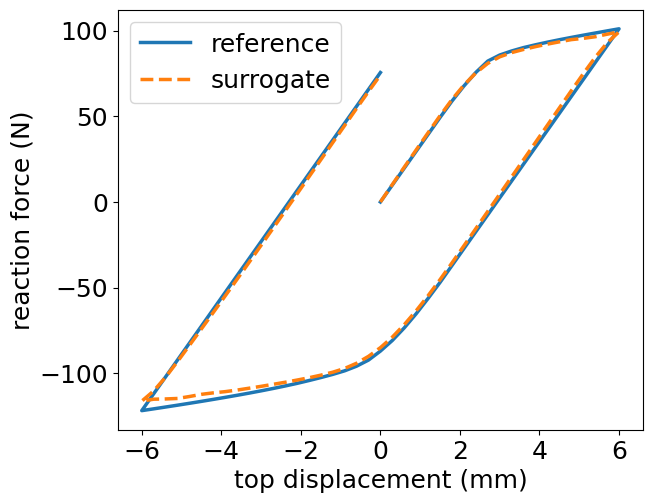}}
    \subfigure[]{\includegraphics[trim =0 0cm 0 0cm, clip, width=7.1cm]{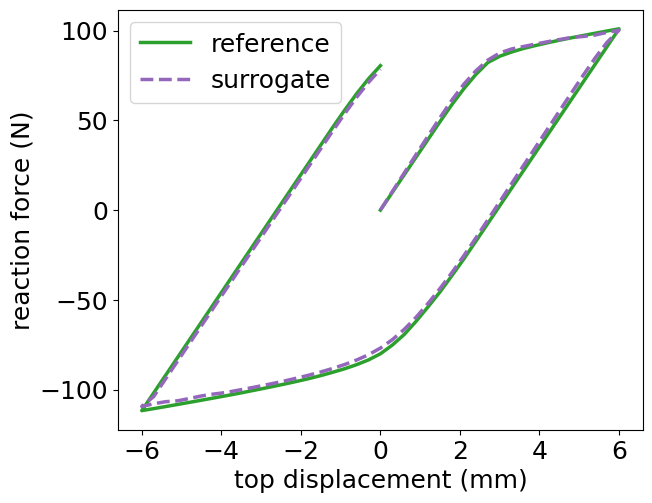}}
    \subfigure[]{\includegraphics[trim =0 0cm 0 0cm, clip, width=7.1cm]{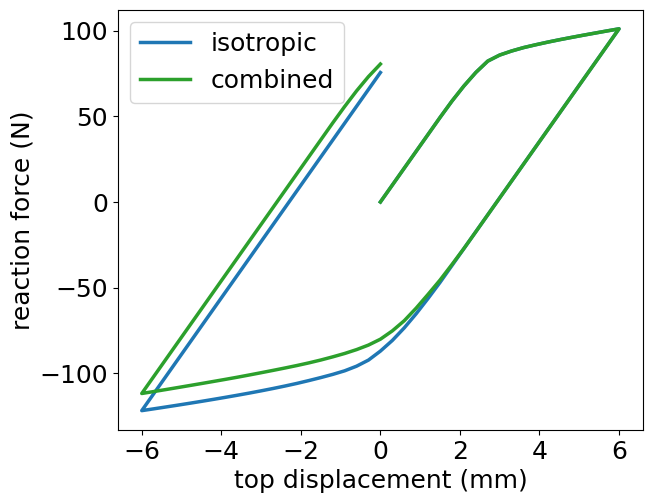}}
    \caption{(a) The loading protocol for the square plate problem; the total reaction force versus controlled displacement for (b) isotropic hardening and (c) combined hardening; (d) comparison between reference solutions solved with isotropic and combined hardening models.}
    \label{quarter_plate_protocol_and_response}
\end{figure}

Consider a $120~\textrm{mm}\times120~\textrm{mm}$ plate with a circular hole in its center (Figure \ref{quarter_plate_bc}a) under the plane strain condition. The radius of the hole is $R=30~\textrm{mm}$.
Due to the symmetry of the problem, only one quarter of the plate is modeled with a structured mesh shown in Figure \ref{quarter_plate_bc}b.
The entire top edge is controlled by a uniform displacement $d=6~\textrm{mm}$ that is scaled  by the loading protocol shown in Figure \ref{quarter_plate_protocol_and_response}a. The bottom of the plate is clamped.  There are three highlighted integration points in Figure \ref{quarter_plate_bc}b which are used to access local stress-strain curves. The element type and material properties are the same as those in Section \ref{sec:coupon}. In particular, this problem is solved for two cases of isotropic and combined hardening ($\hat{\beta}=0.5$) models. For the INCDE model, midpoint method is applied with $\Delta t = 1.0$.

\begin{figure}[ht]
    \centering
    \subfigure[]{\includegraphics[trim =.15cm .5cm .0cm 1cm, clip, width=5cm]{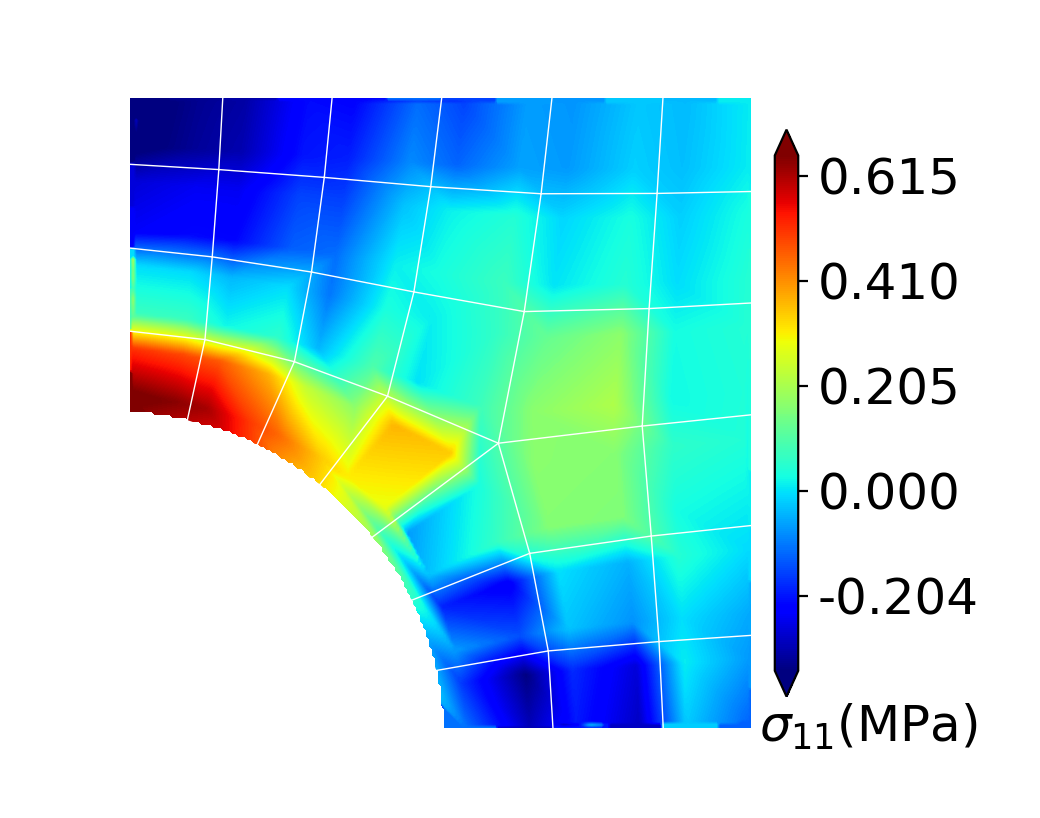}} 
    \subfigure[]{\includegraphics[trim =.15cm .5cm .0cm 1cm, clip, width=5cm]{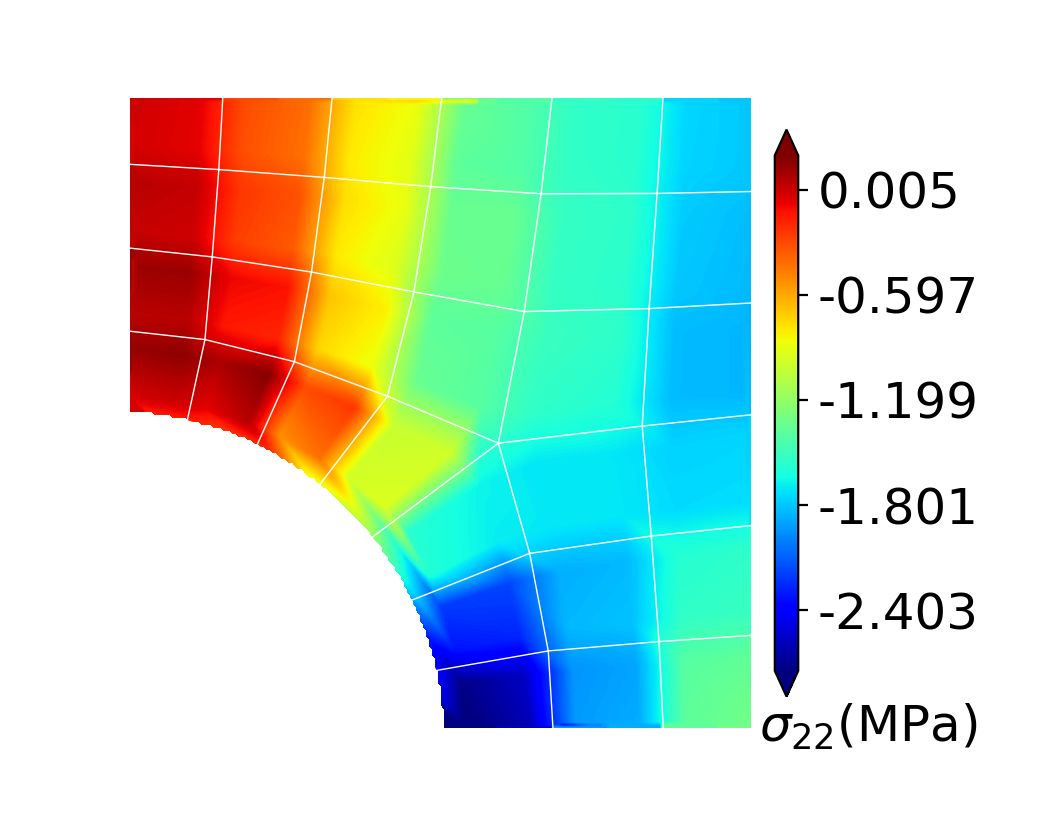}}
    \subfigure[]{\includegraphics[trim =.15cm .5cm .0cm 1cm, clip, width=5cm]{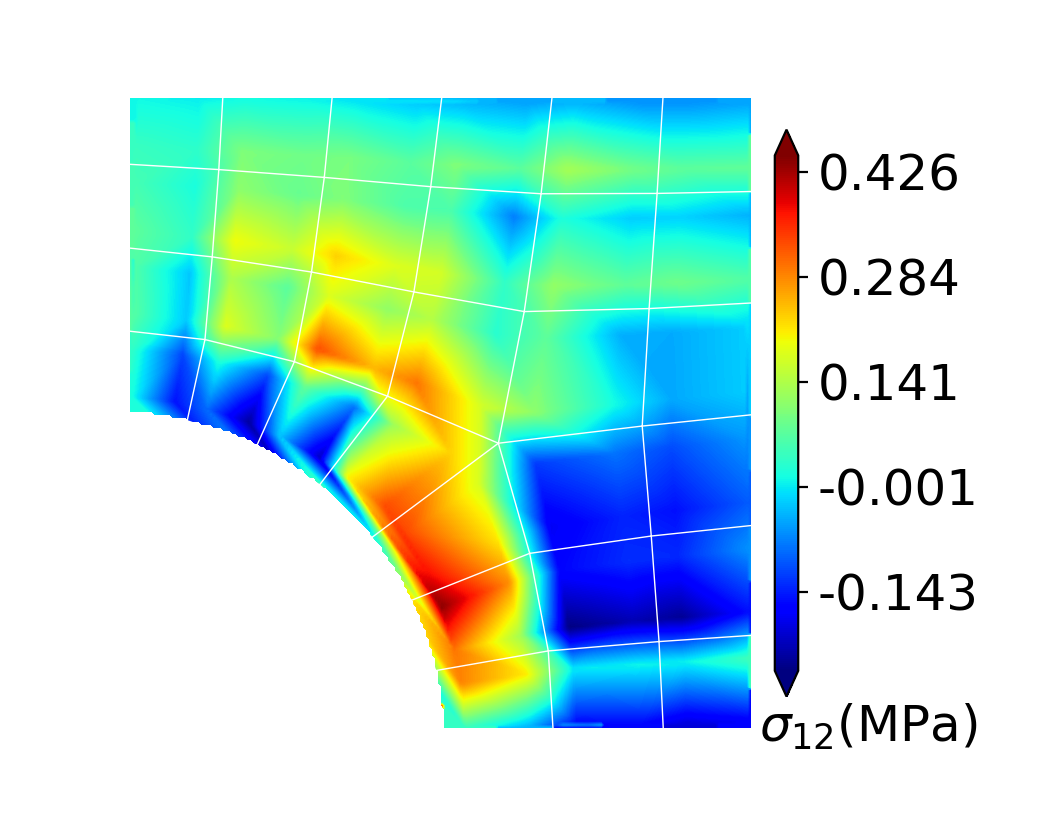}}
    \subfigure[]{\includegraphics[trim =.15cm .5cm .0cm 1cm, clip, width=5cm]{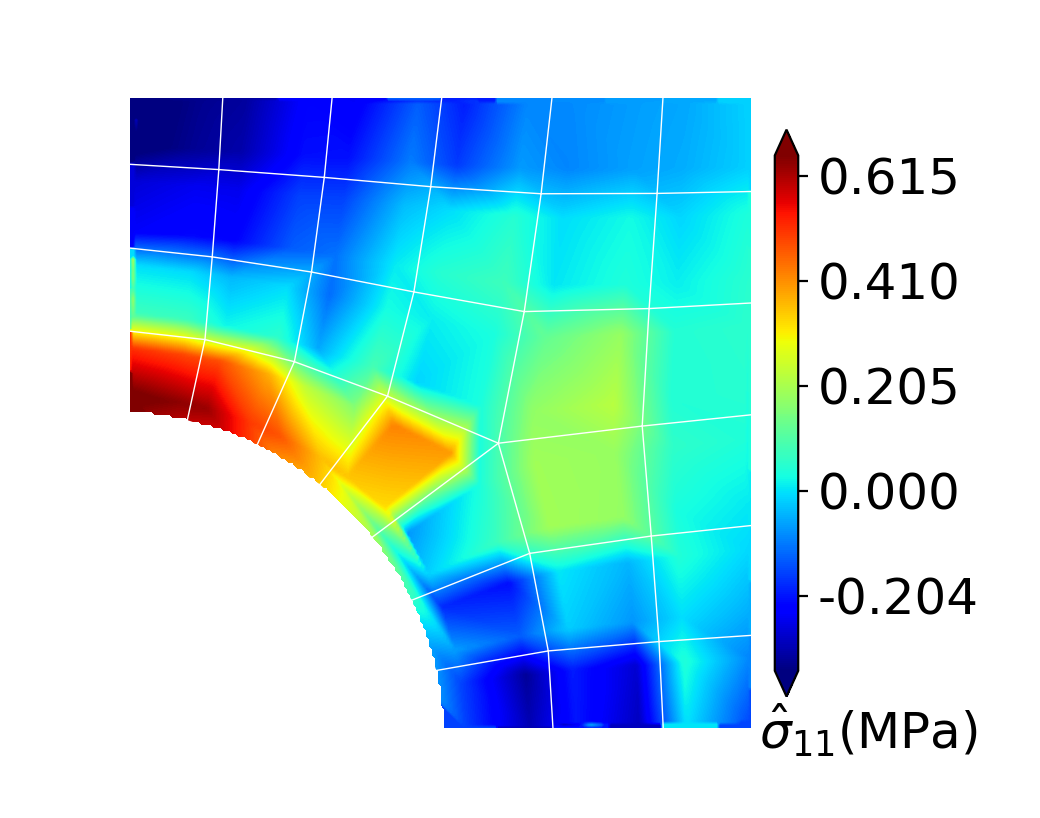}} 
    \subfigure[]{\includegraphics[trim =.15cm .5cm .0cm 1cm, clip, width=5cm]{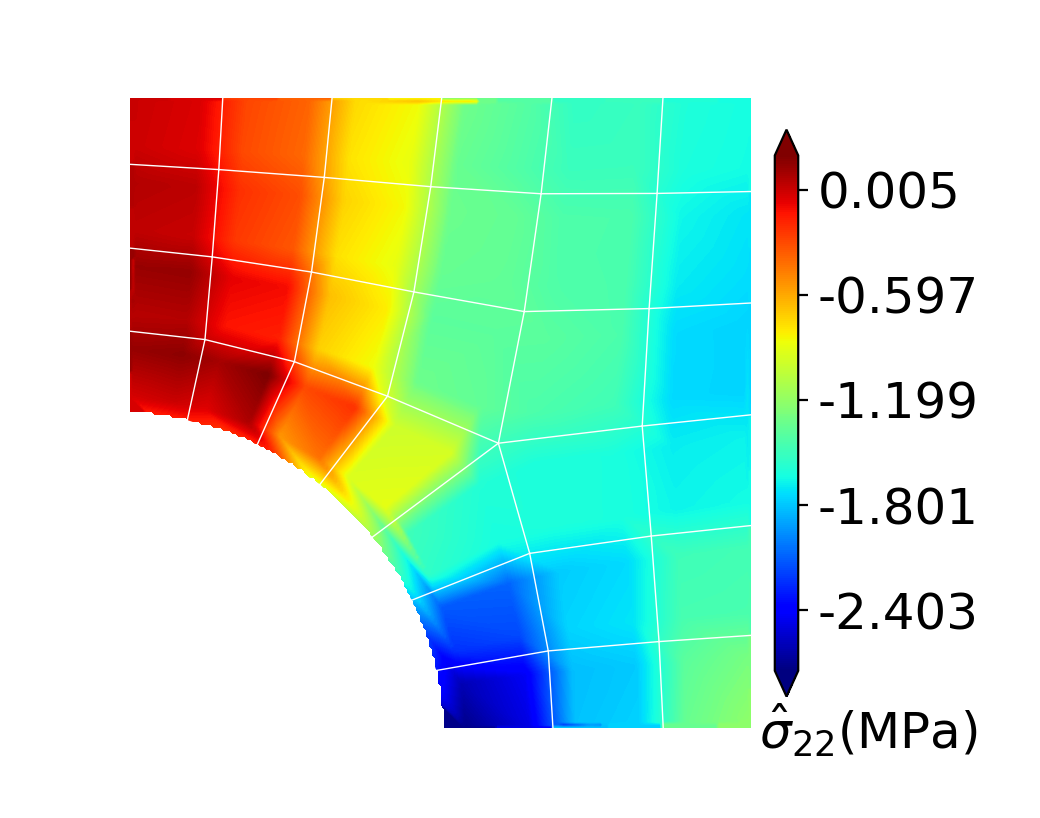}}
    \subfigure[]{\includegraphics[trim =.15cm .5cm .0cm 1cm, clip, width=5cm]{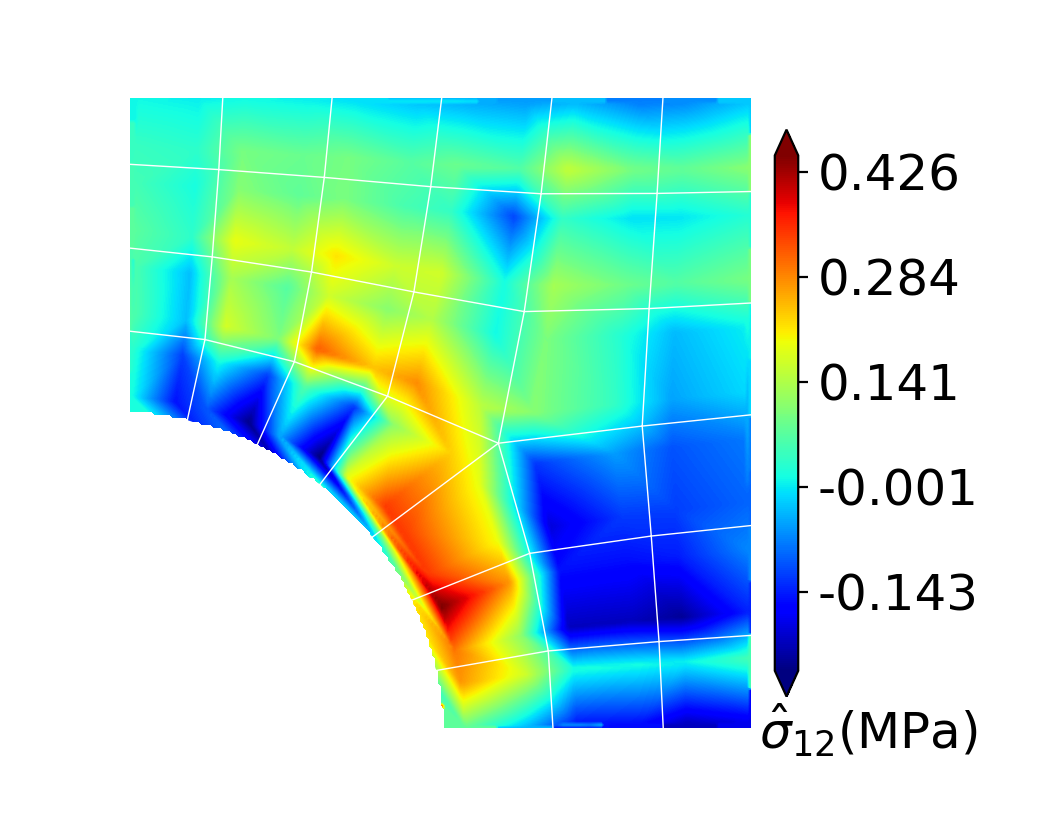}}
    \caption{Stress contour plots of the square plate problem at the minimum loading step ($d = -6~\textrm{mm}$) obtained from conventional FE analysis (first row) and the surrogate model (second row).}
    \label{quarter_plate_linear_stress_min}
\end{figure}

\begin{figure}[ht]
    \centering
    \subfigure[]{\includegraphics[trim =.15cm .5cm .0cm 1cm, clip, width=5cm]{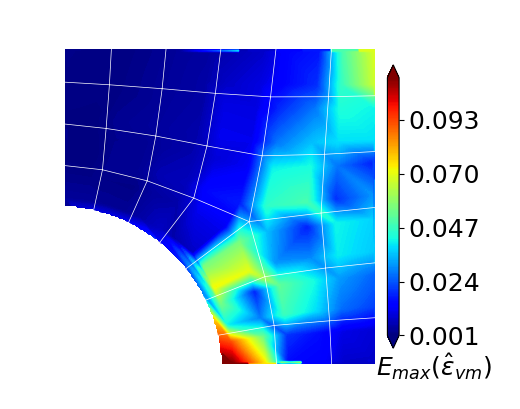}} 
    \subfigure[]{\includegraphics[trim =.15cm .5cm .0cm 1cm, clip, width=5cm]{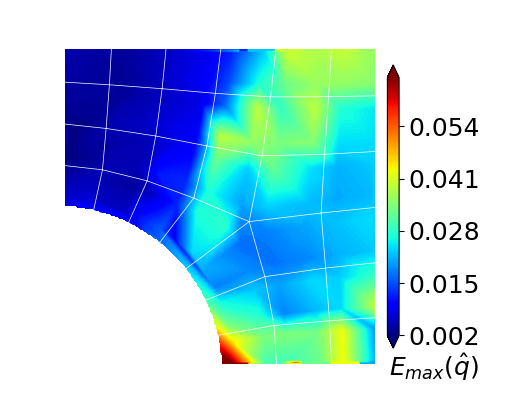}}
    \subfigure[]{\includegraphics[trim =.15cm .5cm .0cm 1cm, clip, width=5cm]{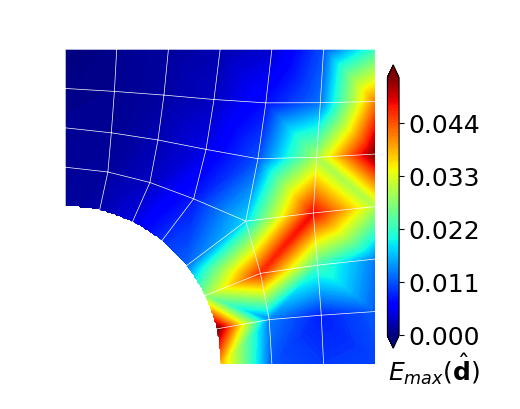}}
    \caption{$E_{max}(\cdot)$ error contour plots of (a) von Mises strain, (b) von Mises stress, and (c) displacement vector for the square plate problem.}
    \label{quarter_plate_err}
\end{figure}

Figures \ref{quarter_plate_protocol_and_response}b and \ref{quarter_plate_protocol_and_response}c show the total reaction force versus the prescribed top displacement for isotropic and combined hardening, respectively. Figure \ref{quarter_plate_protocol_and_response}d is plotted to show differences of the two reference solutions. The overall structural response of the INCDE-based model matches the standard FEM solution well. In addition, Figure \ref{quarter_plate_linear_stress_min} displays a good match between INCDE and standard FEM results for the distribution of all stress components at $d = -6~\textrm{mm}$ in the isotropic hardening case. INCDE successfully replicates the tensile stress $\sigma_{11}$ above the hole and the compressive stress concentration for $\sigma_{22}$ on the right side of the hole. 
Figure \ref{quarter_plate_err} shows the contour error plots (isotropic hardening) for von Mises stress $q$, nodal displacement $\mathbf d$, and von Mises strain $\epsilon_{vm}$ defined as

\begin{equation}\label{vm_strain}
    \begin{split}
        \epsilon_{vm} &:= \sqrt{\frac{2}{3}(e_{11}^2 + e_{22}^2 + e_{33}^2) + \frac{1}{3}(\gamma_{13}^2 + \gamma_{23}^2 + \gamma_{12}^2)}\\
        e_{11} &:= \frac{2}{3}\epsilon_{11} - \frac{1}{3}\epsilon_{22} - \frac{1}{3}\epsilon_{33},\ e_{22} := \frac{2}{3}\epsilon_{22} - \frac{1}{3}\epsilon_{11} - \frac{1}{3}\epsilon_{33} \\ 
        e_{33} &:= \frac{2}{3}\epsilon_{33} - \frac{1}{3}\epsilon_{11} - \frac{1}{3}\epsilon_{22} \\  \, .
    \end{split}
\end{equation}
 Errors are evaluated with the $E_{max}(\cdot)$ error metric defined as 
\begin{equation}\label{Emax_quarter_plate}
    \begin{split}
        E_{max}(\hat{q}_{i}) &= \frac{\Vert \hat{q}_{i} - q_{i} \Vert}{\mathop{\max}_{i\in{S_g}}\Vert q_{i}\Vert}\quad , \quad
        E_{max}(\hat{\epsilon}_{vm, i}) = \frac{\Vert \hat{\epsilon}_{vm, i} - \epsilon_{vm, i} \Vert}{\mathop{\max}_{i\in{S_g}}\Vert\epsilon_{vm, i}\Vert}\\
        E_{max}(\hat{\mathbf d}_i) &= \frac{\Vert \hat{\mathbf d_i} - \mathbf d_i \Vert}{\mathop{\max}_{i\in{S_n}}\Vert\mathbf d_{i}\Vert} \,.
    \end{split}
\end{equation}
In this definition, $q_i$ and $\epsilon_{vm, i}$ are $T\times 1$ arrays for FEM solutions at integration point $i$ and in all time steps from 1 to $T$. $\mathbf d_{i}$ is a $T\times 2$ array for the nodal displacement vector solved with FEM at Node $i$. $S_n$ and $S_g$ represent sets containing all nodes and integration points, respectively. In addition, the quantities with hats denote the corresponding INCDE solutions. This metric measures error of the entire time series at all points normalized with respect to the largest time series norm among all nodes/integration points. This normalization is selected to avoid zero denominators when calculating relative errors. Figure \ref{quarter_plate_err} shows that errors are generally small throughout the domain and mostly do not exceed 5\%. It is noteworthy that the $E_{max}(\cdot)$ metric used here is essentially a normalized measure of the absolute error as opposed to the relative error. Therefore, regions with larger magnitudes of the quantities of interest may be expected to show larger absolute errors.

\begin{figure}[ht]
    \centering
    \includegraphics[trim =0.0cm 2.0cm 0.0cm 1.0cm, clip, width=15cm]{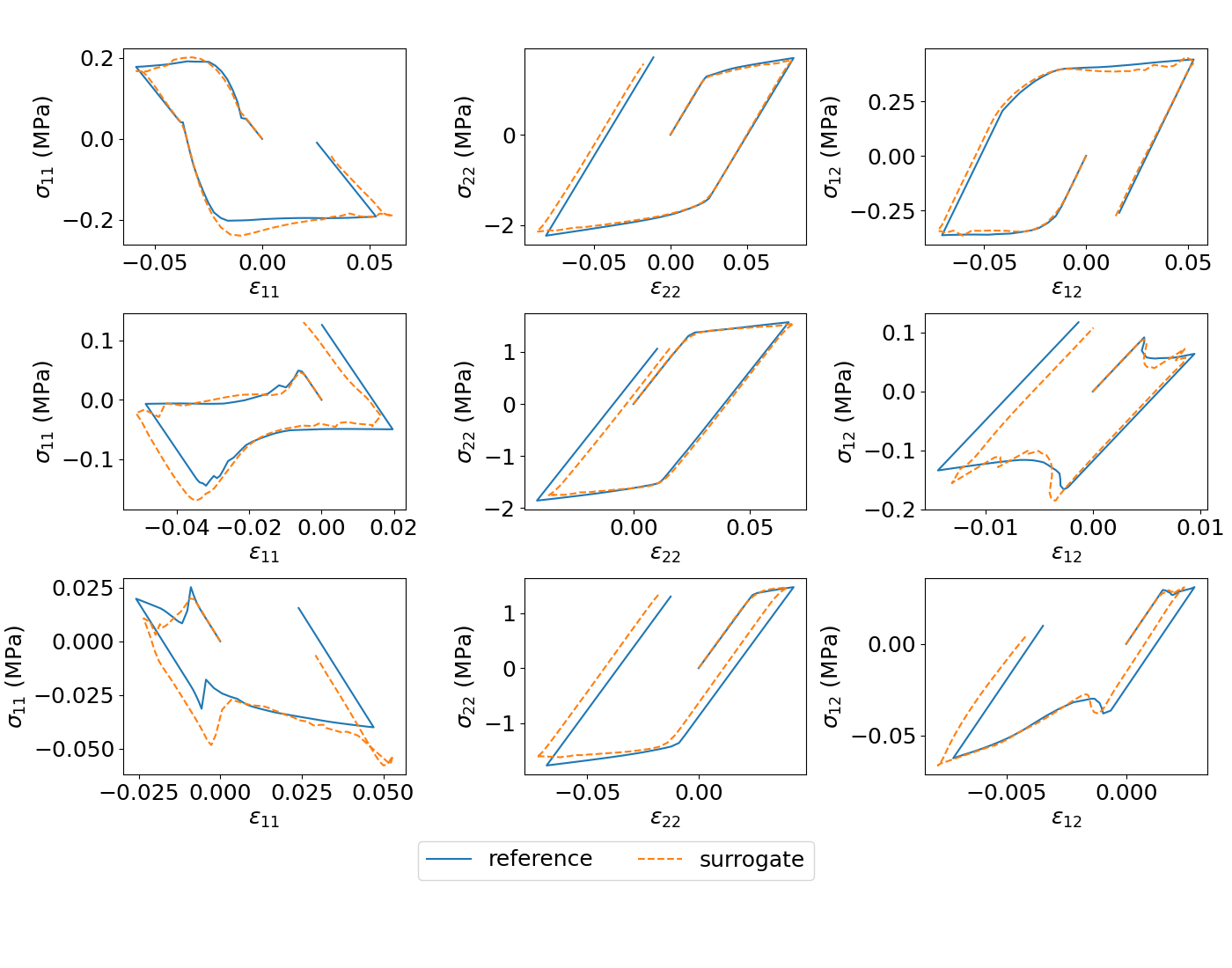}
    \caption{The stress-strain curves at selected integration points shown in Figure \ref{quarter_plate_bc}b. The row numbers match the integration point numbers.}
    \label{quarter_plate_gpt_stress_strain}
\end{figure}

Figure \ref{quarter_plate_gpt_stress_strain} displays stress-strain curves of all components corresponding to the maximum compression stage (i.e. at $d = -6~\textrm{mm}$) for the integration points marked on Figure \ref{quarter_plate_bc}b. It is observed that the INCDE results match very well with those obtained with standard FEM, where generally components with a larger magnitude show a better match. It is noteworthy that the even though the strain patterns resulted from loading in this problem were not included in the partitioned random-walk-based training dataset, the INCDE-based surrogate model has captured the overall behavior accurately.

\subsubsection{Shear Localization}
\begin{figure}[ht]
    \centering
    \subfigure[]{\includegraphics[trim =0cm .75cm 0cm 1.cm, clip, width=7cm]{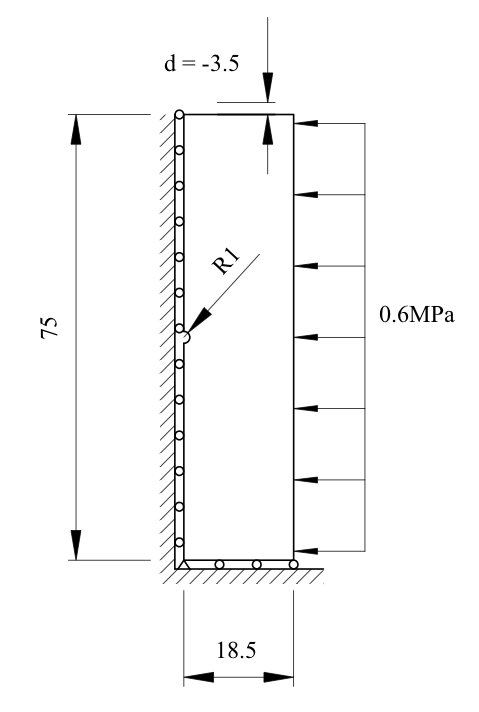}}
    \subfigure[]{\includegraphics[trim =0cm .0cm 0cm 0.cm, clip, width=8cm]{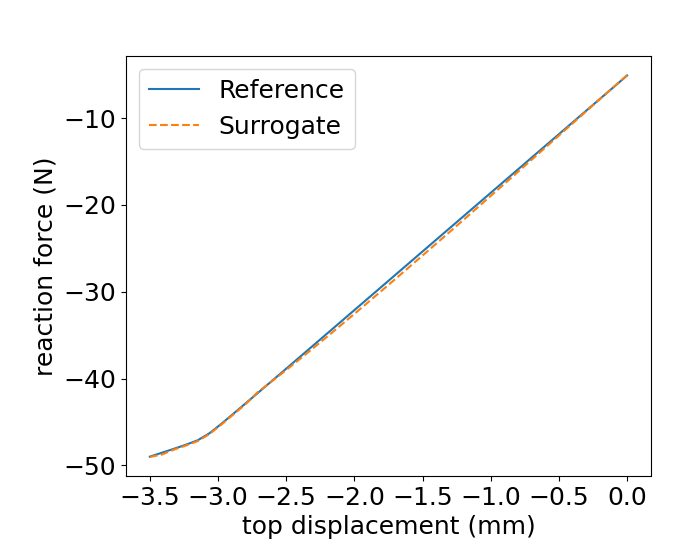}}
    \caption{(a) The geometry and boundary conditions of the shear localization problem; and (b) the reaction force versus controlled top displacement in the second stage of loading.}
    \label{shear_loc_geo_mesh}
\end{figure}

This example aims to test the capability of the INCDE model in capturing shear localization. A specimen shown in Figure \ref{shear_loc_geo_mesh}a with a small circular imperfection in its center is considered under the plane strain condition. Boundary conditions are applied in two stages. First, a uniform pressure of $p = 0.6$ MPa is exerted to the hole and lateral edges. Subsequently, the top edge of the specimen is controlled by prescribed displacement, decreasing to $-3.5$mm uniformly in 40 steps. In both stages, the bottom of the specimen is fixed, and symmetric boundary conditions are applied along the vertical axis. The material model is Drucker-Prager with same properties used in the dataset described in Section \ref{sec:data}.

\begin{figure}[ht]
    \centering
    \subfigure[]{\includegraphics[trim =0cm 1.5cm 0cm 1.5cm, clip, width=3.1cm]{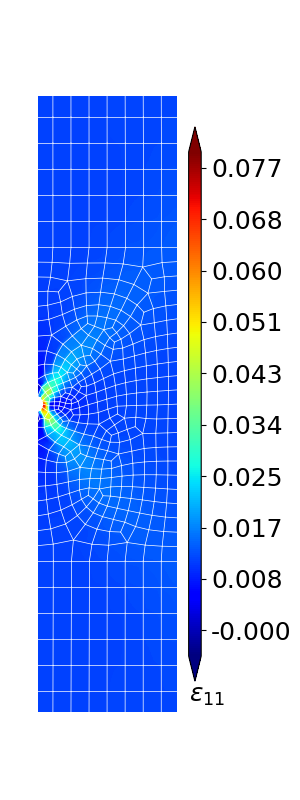}} 
    \subfigure[]{\includegraphics[trim =0cm 1.5cm 0cm 1.5cm, clip, width=3.1cm]{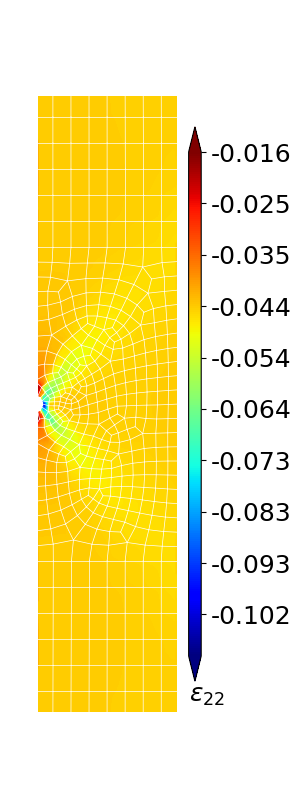}} 
    \subfigure[]{\includegraphics[trim =0cm 1.5cm 0cm 1.5cm, clip, width=3.1cm]{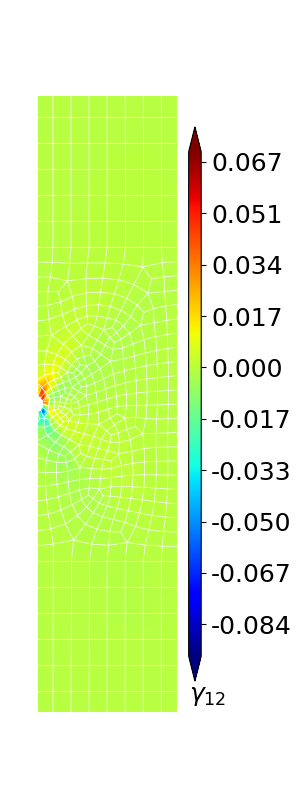}}\\
    \subfigure[]{\includegraphics[trim =0cm 1.5cm 0cm 1.5cm, clip, width=3.1cm]{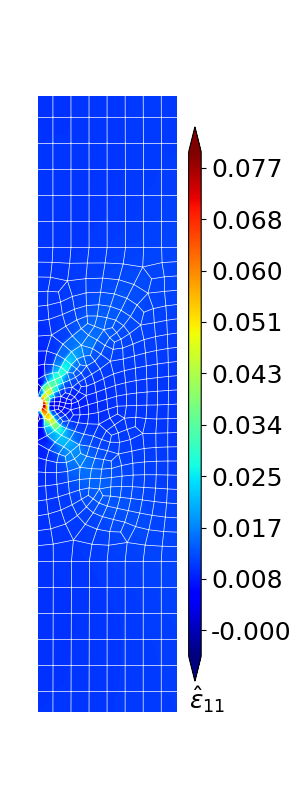}}
    \subfigure[]{\includegraphics[trim =0cm 1.5cm 0cm 1.5cm, clip, width=3.1cm]{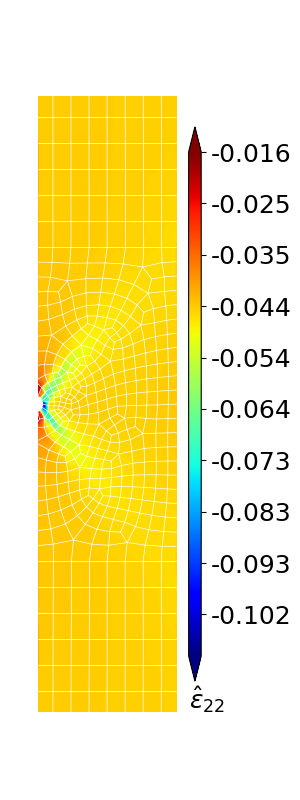}}
    \subfigure[]{\includegraphics[trim =0cm 1.5cm 0cm 1.5cm, clip, width=3.1cm]{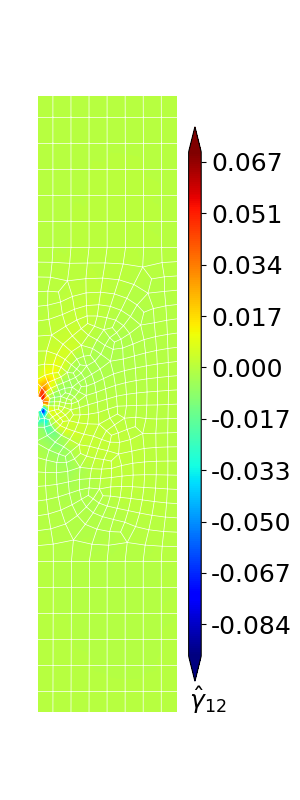}}
    \caption{Contour plots of all strain components from FE analyses with the traditional constitutive (a-c) and surrogate (d-f) models in the final loading step.}
    \label{shear_loc_contourf_strain}
\end{figure}

\begin{figure}[ht]
    \centering
    \subfigure[]{\includegraphics[trim =0cm 1.5cm 0cm 1.5cm, clip, width=3.1cm]{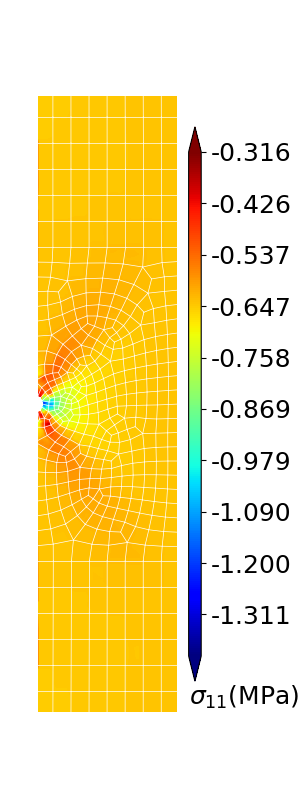}} 
    \subfigure[]{\includegraphics[trim =0cm 1.5cm 0cm 1.5cm, clip, width=3.1cm]{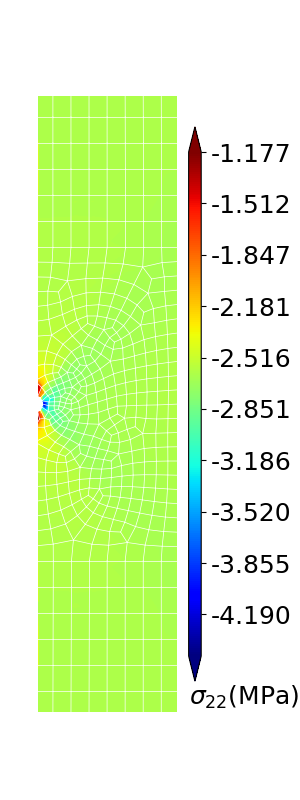}} 
    \subfigure[]{\includegraphics[trim =0cm 1.5cm 0cm 1.5cm, clip, width=3.1cm]{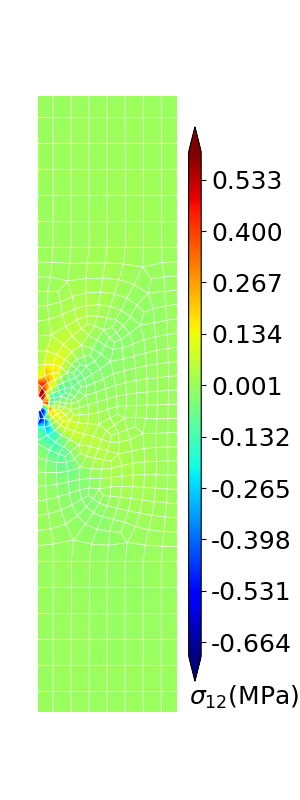}}\\
    \subfigure[]{\includegraphics[trim =0cm 1.5cm 0cm 1.5cm, clip, width=3.1cm]{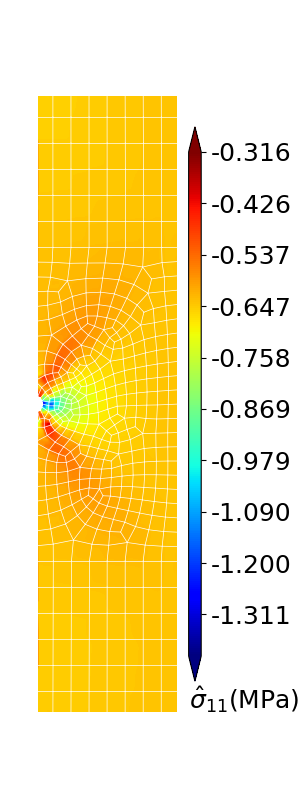}}
    \subfigure[]{\includegraphics[trim =0cm 1.5cm 0cm 1.5cm, clip, width=3.1cm]{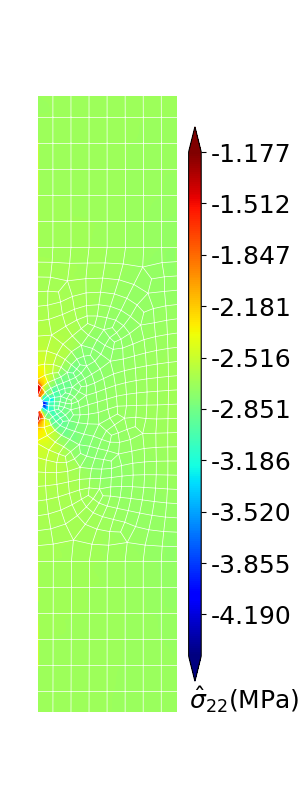}}
    \subfigure[]{\includegraphics[trim =0cm 1.5cm 0cm 1.5cm, clip, width=3.1cm]{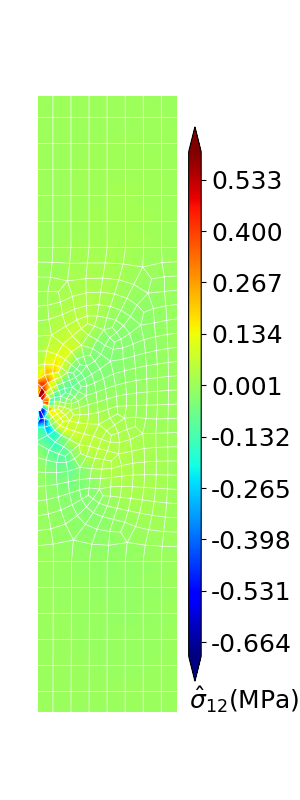}}
    \caption{Contour plots of all stress components from FE analyses with the traditional constitutive (a-c) and surrogate (d-f) models in the final loading step.}
    \label{shear_loc_contourf_stress}
\end{figure}

\begin{figure}[ht]
    \centering
    \subfigure[]{\includegraphics[trim =0cm 1.5cm 0cm 1.5cm, clip, width=3.1cm]{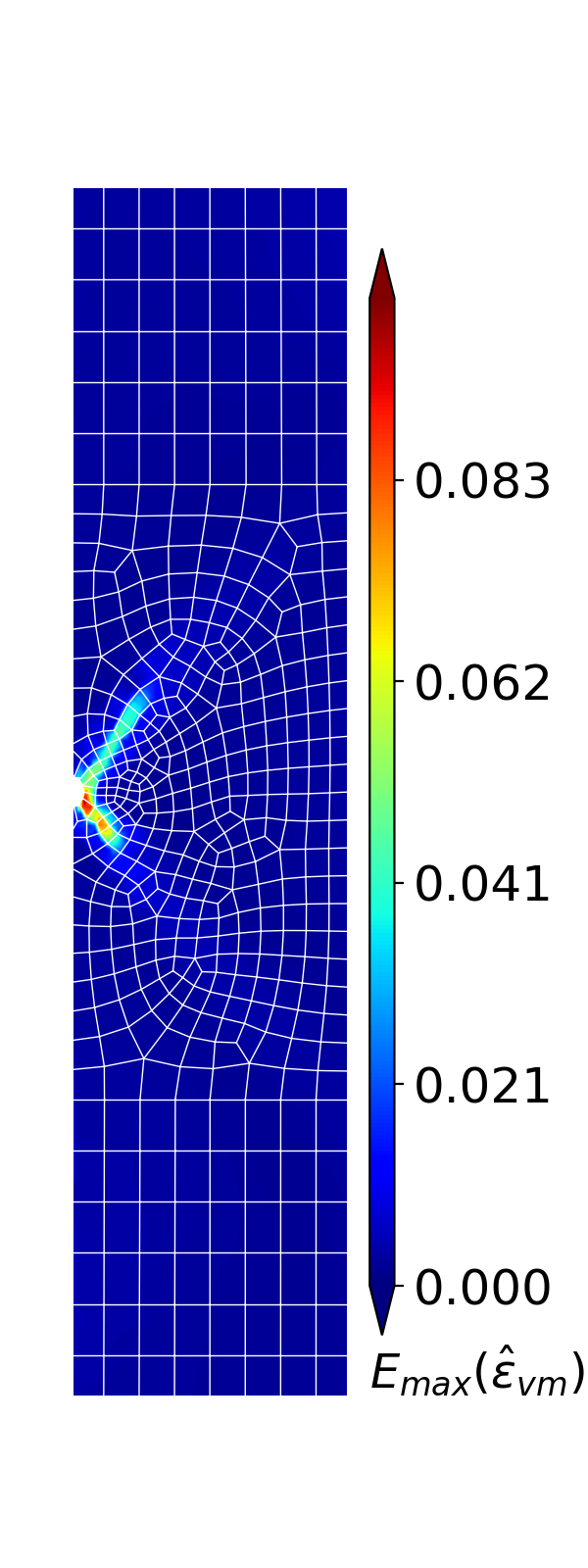}} 
    \subfigure[]{\includegraphics[trim =0cm 1.5cm 0cm 1.5cm, clip, width=3.1cm]{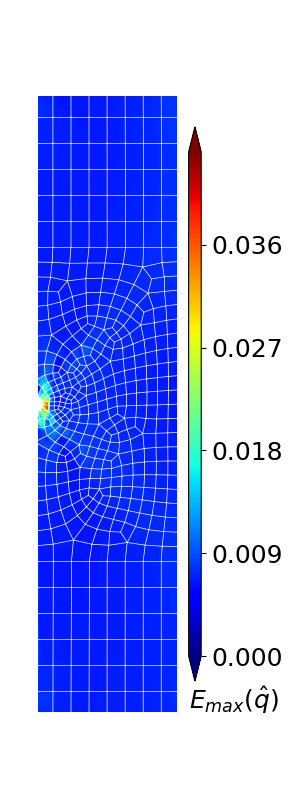}} 
    \subfigure[]{\includegraphics[trim =0cm 1.5cm 0cm 1.5cm, clip, width=3.1cm]{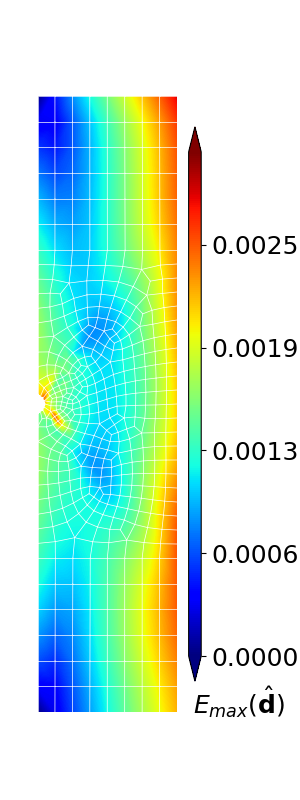}}\\
    \subfigure[]{\includegraphics[trim =0cm 1.5cm 0cm 1.5cm, clip, width=3.1cm]{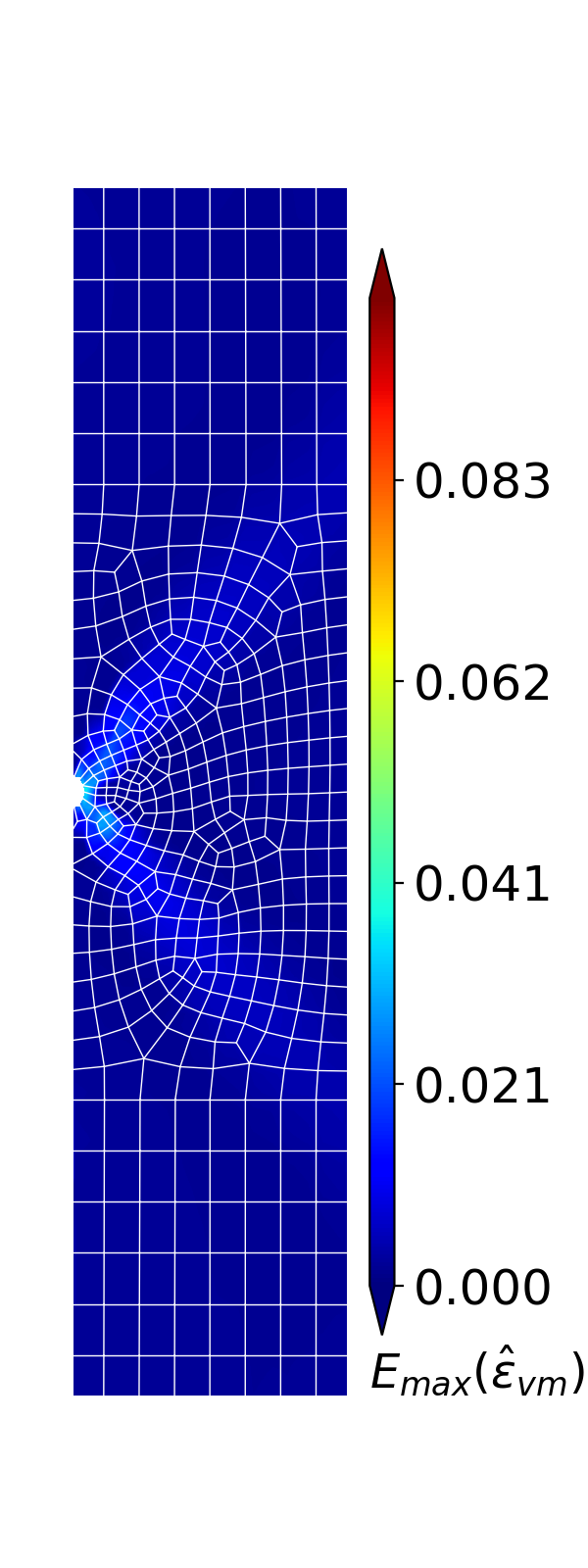}} 
    \subfigure[]{\includegraphics[trim =0cm 1.5cm 0cm 1.5cm, clip, width=3.1cm]{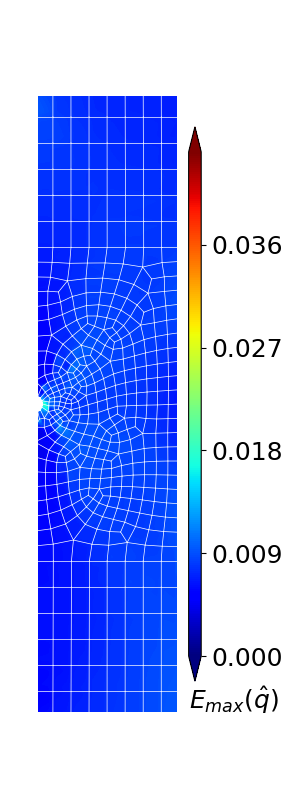}} 
    \subfigure[]{\includegraphics[trim =0cm 1.5cm 0cm 1.5cm, clip, width=3.1cm]{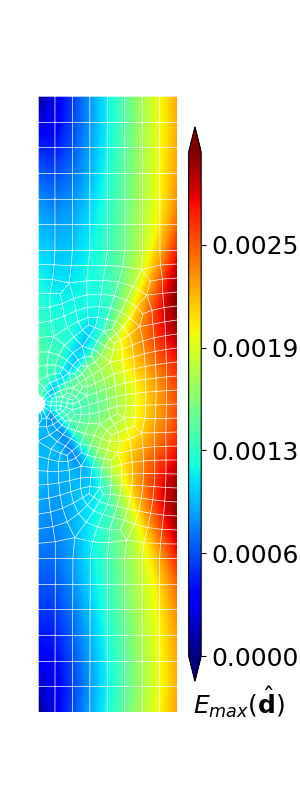}}
    \caption{$E_{max}(\cdot)$ error contour plots of von Mises strain, von Mises stress, and displacement corresponding to the INCDE model without (a-c) and with (d-e) volumetric-deviatoric decomposition.}
    \label{shear_loc_error_contourf}
\end{figure}

Figure \ref{shear_loc_geo_mesh}b shows the reaction force on the top edge versus controlled displacement during the second loading stage. It is observed that the global structural response in both elastic and elasto-plastic stages are predicted well. The global yielding point is also predicted accurately. Figures \ref{shear_loc_contourf_strain} and \ref{shear_loc_contourf_stress} provide strain and stress contours at the end of loading (i.e., $d = -3.5$mm), respectively. We observe that the surrogate model predicts the shear band and stress concentration near the imperfection very well, with the magnitude of the concentrated stress and the shear band angle matching those from the reference solution. In addition, Figure \ref{shear_loc_error_contourf} visualizes the error distribution of the von Mises strain, von Mises stress, and displacement of INCDE models trained with and without volumetric-deviatoric decomposition of stress. It is observed that the INCDE model trained to output pressure and deviatoric stress components has much smaller errors near the circular hole and shear bands compared to the INCDE model trained without stress decomposition. The model trained with stress decomposition results in errors smaller than 4\% and 1\% for von Mises strain and stress, respectively, which significantly outperforms the model trained without decomposition. The displacement error magnitudes in these two cases are similar and bounded by 0.3\%. Therefore, it is concluded that stress decomposition can improve the accuracy of the INCDE model for Drucker-Prager plasticity.

\section{Conclusion}

This work shows the robustness of neural differential equations in modeling elasto-plastic path-dependent material behavior. We have designed, analyzed, and implemented Incremental Neural Controlled Differential Equations (INCDE), a new machine learning model for time-variant systems based on an extension of Neural Controlled Differential Equations (NCDE). INCDE was applied to surrogate modeling of path-dependent material behavior for implementation within standard nonlinear FE solvers. 

We first formulated the rate-form of INCDE as a nonlinear time-variant system for the internal state variable (hidden state), in which a neural network represents the modulus matrix, similar to NCDE. The formulation is further enhanced by adding the controlling strain sequence and its rate as inputs of this neural network to improve approximation of path-dependent material behavior with irregular strain increments. 
Subsequently, the incremental form of INCDE was derived by representing the strain sequence as a piecewise linear function. As a result, a neural initial value problem is defined in each load step and is initialized using the hidden state calculated in the previous step. Finally, the dynamical system was stabilized with a self-damping factor.

Advantages of the new INCDE formulation include flexibility in capturing functions with low and high degrees of continuity, avoiding the need for interpolation of time series before integration, preserving consistency of the hidden state with respect to the control strain sequence, self-consistency of the stress predictions with respect to strain increments, applicability to irregular time/strain series,  as well as stability and boundedness of the predictions. 
Using the proposed INCDE model, we developed a surrogate model for stress prediction trained with a random-walk-based dataset. 
Simulations of single integration points as well as boundary value problems were conducted under various loading protocols to evaluate the performance of the trained model. 
Using a series of loading protocols with varying step sizes, we observed that the model captures time history and spatial variations of field variables very well for various loading scenarios and patterns which were not included in the training dataset. 

The convergence properties of the INCDE-based model were validated theoretically and numerically using different ODE solvers. It was demonstrated that the predicted stresses and state variables show linear convergence with strain increment refinement, and their convergence rate with respect to time step size agrees with the theoretical convergence rate of the ODE solvers. 
While the hidden state and stress are shown to converge to a certain limit, it is noted that this limit may not necessarily match the ground truth, and it is the main goal of INCDE training to map this numerical solution to the ground truth. The strategies implemented in this work for this purpose include training INCDE with small step sizes and higher-order solvers (e.g. RK4 and midpoint method), and building a training dataset with small time step sizes based on the random-walk method. Design of the training datasets and training methods are critical factors in maximizing the usage of the convergence property of INCDE-based models. 

Examples of boundary value problems in which the INCDE surrogate model replaced the conventional constitutive models demonstrated the capability of the INCDE to generate a seamless and accurate global structural response. Overall, our findings show the potential of INCDE-based model as a promising strategy for robust modeling of the behavior of complex materials by taking advantage of the properties of ordinary differential equations in capturing time-dependent systems along with the flexibility of neural networks.

\section*{Acknowledgement}
This material is based upon research supported by the Office of Naval Research under Award Number N00014-23-1-2180.

\end{document}